\documentclass[
 reprint,
superscriptaddress,
 amsmath,amssymb,
 aps,journal, floatfix, physrev, nofootinbib
]{revtex4-2}

\usepackage{graphicx}
\usepackage{dcolumn}
\usepackage{bm}
\usepackage{url} 
\usepackage{bm}
\usepackage{hyperref}
\usepackage{subcaption}

\usepackage{caption}
\begin{document}


\title{Hyper Suprime-Cam Y3 results: photo-$z$ bias calibration with lensing shear ratios and cosmological constraints from cosmic shear}

\author{Divya Rana}
\email{rana@strw.leidenuniv.nl}
\affiliation{Leiden Observatory, Leiden University, PO Box 9513, NL-2300 RA Leiden, The Netherlands}
\affiliation{Inter University Centre for Astronomy and Astrophysics, Ganeshkhind, Pune 411007, India}
\author{Surhud More}%
\email{surhud@iucaa.in}
\affiliation{Inter University Centre for Astronomy and Astrophysics, Ganeshkhind, Pune 411007, India}
\affiliation{Kavli Institute for the Physics and Mathematics of the Universe (WPI), University of Tokyo, 5-1-5, Kashiwanoha, 2778583, Japan}
\author{Hironao Miyatake}%
\affiliation{Kobayashi-Maskawa Institute for the Origin of Particles and the Universe (KMI), Nagoya University, Furo-cho, Chikusa-ward, Nagoya, Aichi, 464-8602, Japan}
\affiliation{Institute for Advanced Research, Nagoya University, Furo-cho, Chikusa-ward, Nagoya, Aichi, 464-8601, Japan}
\affiliation{Kavli Institute for the Physics and Mathematics of the Universe (WPI), University of Tokyo, 5-1-5, Kashiwanoha, 2778583, Japan}

\author{Sunao Sugiyama}
\affiliation{Department of Physics and Astronomy, University of Pennsylvania, Philadelphia, PA 19104, USA}

\author{Tianqing Zhang}%
\affiliation{Department of Physics and Astronomy and PITT PACC, University of Pittsburgh, Pittsburgh, PA 15260, USA}

\author{Masato Shirasaki}%
\affiliation{National Astronomical Observatory of Japan (NAOJ), National Institutes of Natural Science, Mitaka, Tokyo 181-8588, Japan}
\affiliation{The Institute of Statistical Mathematics, Tachikawa, Tokyo 190-8562, Japan}


\date{\today}

\begin{abstract}
We present an independent calibration of the photometric redshift (photo-$z$) distributions for source galaxies in the HSC-Y3 weak lensing survey using small-scale galaxy-galaxy lensing. By measuring the tangential shear around spectroscopic lens galaxies from GAMA, SDSS, and DESI, divided into fifteen narrow redshift bins, we compute shear ratios that are sensitive to the mean redshift of source galaxies. Using a blinded analysis, we derive constraints on the photo-$z$ bias parameters in source bins 2, 3 and 4, achieving signal-to-noise ratios of 59, 75, and 62, respectively. Our constraints for $\Delta z_2$, $\Delta z_3$ and $\Delta z_4$ are consistent with those from HSC-Y3 cosmic shear modeling. We observe a mild shift in the $\Delta z_3$--$\Delta z_4$ plane due to the heterogeneous depth of the lens sample, which disappears when using only DESI-DR1 lenses. Combining shear-ratio measurements with cosmic shear data, we obtain joint constraints on cosmological parameters: $\Omega_{\rm m} = 0.304_{-0.029}^{+0.03}$ and $S_8 = 0.773_{-0.031}^{+0.031}$, consistent with cosmic shear-only results. This work demonstrates the utility of small-scale lensing as an independent probe for calibrating photometric redshift bias in weak lensing cosmology.
\end{abstract}

\maketitle


\section{Introduction}

When the light from distant galaxies traverses the Universe, it is subtly deflected by the gravitational influence of intervening matter via a phenomenon known as \emph{weak gravitational lensing}~\citep{1937_Zwicky,1992_Kaiser}. Unlike strong lensing, which produces dramatic features such as arcs and multiple images, weak lensing induces small, coherent distortions in the observed shapes of background galaxies. These distortions, although individually faint, can be statistically measured over large samples to infer the projected matter distribution along the line of sight~\citep[see, e.g.,][]{1992_Kaiser,2001_Bartelmann,2008_Hoekstra,2015_Kilbinger,2018_Mandelbaum}. A particularly informative application is \emph{galaxy-galaxy lensing}, which correlates the positions of foreground (lens) galaxies with the ellipticities of background (source) galaxies. This cross-correlation probes the matter density around lens galaxies and provides insight into the galaxy-matter connection~\citep[see e.g.,][]{2006_Mandelbaum,2012_Leauthaud,2016_Van_uitert,2013_Cacciato,2015_More,2022_Rana,2023_Mishra}.

Interpreting galaxy--galaxy lensing signals, however, presents modeling challenges. On small angular scales, where signal-to-noise is high, the signal becomes sensitive to complex effects such as non-linear structure formation, baryonic feedback, and the scale-dependent nature of galaxy bias~\citep{1993_Fry,2011_Van_daalen,2013_Van_den_bosch,2013_Semboloni,2015_Harnois}. These factors introduce significant theoretical uncertainties and typically limit the use of small-scale information in standard cosmological analyses.

To address these limitations, several studies have proposed using \emph{lensing ratios}, defined as ratios of galaxy--shear correlation functions measured using a common lens sample but different source populations. These ratios preserve sensitivity to angular diameter distance ratios while largely canceling dependence on the small-scale matter power spectrum and galaxy bias. Consequently, they offer a robust geometrical probe of cosmology that is less affected by uncertainties in small-scale modeling.

The lensing ratio method was first proposed by \citet{2003_Jain} as a way to isolate geometric information and constrain the dark energy equation-of-state parameter \(w\). \citet{2007_Taylor} extended this approach to individual shear measurements behind galaxy clusters, with early observational implementations by \citet{2007_Kitching} and further refinements using data from the HST COSMOS survey~\citep{2012_Taylor}. Later work expanded the method to include both galaxy--shear and shear--shear correlations~\citep{2005_Zhang}, while \citet{2004_Bernstein} developed an alternative formalism and emphasized the need for precise photometric redshifts and shear calibration. \citet{2008_Kitching} further highlighted that photometric redshift errors often dominate the uncertainty in lensing ratio measurements.

Due to their sensitivity to redshift systematics, lensing ratios derived from galaxy-galaxy lensing are particularly useful for testing source redshift distributions and redshift-dependent shear calibration biases~\citep{2016_Schneider,2024_Emas}. Although cosmological constraints can be extracted from these ratios—especially when combined with external datasets such as CMB lensing~\citep{2009_Das,2015_Kitching,2017_Miyatake,2019_Prat}—their constraining power on their own is often limited by uncertainties in redshift estimation. As a result, many surveys utilize the \emph{shear-ratio test} as a consistency check for photometric redshift calibration. This test involves comparing lensing signals across different source redshift bins for a fixed lens bin to validate the inferred redshift distributions. It has been applied in several major wide-field surveys, including SDSS~\citep{2005_Mandelbaum}, RCS~\citep{2005_Hoekstra}, KiDS~\citep{2012_Heymans,2017_Hildebrandt,2020_Hildebrandt,2021_Giblin}, and DES~\citep{2018_Prat,2022_Sanchez}.

In this work, we apply the lensing ratio method to the Year 3 data from the Hyper Suprime-Cam (HSC) survey~\citep{2018_Aihara}. We construct ratios of tangential shear measurements across multiple source redshift bins, all sharing a common spectroscopic lens redshift bin. These ratios allow us to leverage small-scale lensing signals, which are typically excluded from standard two-point analyses due to modeling uncertainties. The method effectively suppresses dependence on quantities such as galaxy bias, the non-linear matter power spectrum, and baryonic effects. As a result, the signal depends primarily on geometry, allowing for a clean and interpretable theoretical prediction.

Our analysis adopts the same cosmological modeling framework as the HSC Y3 cosmic shear analysis~\citep{2023_Li_xianchong,2023_Dalal}, using consistent treatments of nuisance parameters such as intrinsic alignments, source redshift distributions, and shear calibration biases. We implement a blinding strategy in the model space by perturbing the input source redshift distributions. We first derive constraints on photometric redshift (photo-\textit{z}) systematics using the shear-ratio measurements and compare these results with those obtained from modeling the cosmic shear data. We then incorporate the shear-ratio measurements into a joint analysis with cosmic shear and assess their impact on cosmological parameter constraints. We find that the shear-ratio measurements provide complementary information and remain consistent with the cosmic shear-only results for both photo-\textit{z} systematics and cosmological parameters. The full methodology and baseline cosmic shear analysis are detailed in the HSC-Y3 cosmological analysis papers~\citep{2023_Li_xianchong,2023_Dalal,2023_More,2023_Sugiyama,2023_Miyatake}.

The structure of this paper is as follows. In Sec.~\ref{sec:data}, we describe the observational data sets used in our analysis. The methodology for measuring the lensing ratios from the data is presented in Sec.~\ref{sec:meas}, followed by a description of the theoretical model in Sec.~\ref{sec:model} and a discussion of the associated systematics in Sec.~\ref{sec:systematics}. In Sec.~\ref{sec:bayesian}, we outline our Bayesian inference framework for constraining the model parameters, using the measurements described in Secs.~\ref{sec:shear} and \ref{sec:shear-ratio}. In Sec.~\ref{sec:fiducial-constraints}, we evaluate the constraining power of the lensing ratios, both on their own and in combination with other cosmological probes. We summarize our findings and present our conclusions in Sec.~\ref{sec:summary}.

\section{Data}\label{sec:data}
Here we describe the lens and source galaxy samples used in this analysis. The background source galaxies are drawn from the HSC-Y3 shape catalog \citep{2022_Li_xianchong}, which has been extensively utilized in the HSC-Y3 cosmology analysis \citep{2023_More, 2023_Li_xianchong, 2023_Miyatake, 2023_Dalal, 2023_Sugiyama}.

\subsection{Lenses}
\begin{table}
\setlength{\tabcolsep}{2pt}
\begin{center}
\begin{tabular}{cccc}  \hline\hline
{ Bin no} & {Lens bin} & { Source bin $z_{\rm src}$} & { Number of lenses} \\ \hline
$1$  & 0.04$<z_{\rm l}<$0.08 & 2,3,4 & 13971 \\ 
$2$  & 0.08$<z_{\rm l}<$0.12 & 2,3,4 & 19692 \\ 
$3$  & 0.12$<z_{\rm l}<$0.16 & 2,3,4 & 26428 \\ 
$4$  & 0.16$<z_{\rm l}<$0.20 & 2,3,4 & 25181 \\ 
$5$  & 0.20$<z_{\rm l}<$0.24 & 2,3,4 & 23546 \\ 
$6$  & 0.24$<z_{\rm l}<$0.28 & 2,3,4 & 23979 \\ 
$7$  & 0.28$<z_{\rm l}<$0.32 & 2,3,4 & 21258 \\ 
$8$  & 0.32$<z_{\rm l}<$0.36 & 2,3,4 & 14202 \\ 
$9$  & 0.36$<z_{\rm l}<$0.40 & 2,3,4 & 9827 \\ 
$10$  & 0.11$<z_{\rm l}<$0.17 & 3,4   & 34576 \\ 
${11}$ & 0.17$<z_{\rm l}<$0.22 & 3,4   & 35173 \\ 
${12}$ & 0.22$<z_{\rm l}<$0.28 & 3,4   & 32225 \\ 
${13}$ & 0.28$<z_{\rm l}<$0.33 & 3,4   & 28229 \\ 
${14}$ & 0.33$<z_{\rm l}<$0.39 & 3,4   & 16455 \\ 
${15}$ & 0.39$<z_{\rm l}<$0.44 & 3,4   & 10321 \\ 
${16}$ & 0.44$<z_{\rm l}<$0.50 & 3,4   & 9360 \\ 
\hline
\end{tabular}
\end{center}
\caption{
Lens samples used in the analysis, categorized by lens redshift bin \( z_l \), along with corresponding source redshift bins and the number of lenses used for weak lensing signal measurements.}
\label{tab:lenses}
\end{table}

\begin{figure}
    \centering
    \includegraphics[width=\linewidth]{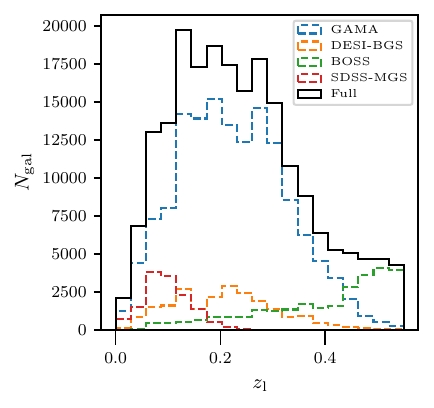}
    \caption{{\it Lens redshift distribution}: The figure shows the lens redshift distribution of our fiducial sample. The solid black line denotes the distribution for the full sample. The blue, orange, green, and red dashed lines represent the redshift distributions of the individual GAMA, DESI-BGS, BOSS, and SDSS-MGS samples, respectively.
}
    \label{fig:lens-redshift}
\end{figure}

We construct our lens sample using spectroscopic galaxy data from multiple surveys overlapping with the HSC-Y3 footprint to maximize the signal-to-noise ratio of the weak lensing shear measurements. In regions where surveys overlap, we retain lenses from the deeper spectroscopic dataset following the priority order: GAMA, DESI, BOSS, and SDSS-MGS. For surveys of similar depth, such as BOSS and SDSS-MGS, we remove duplicate galaxies to avoid double counting. The combined redshift distribution of all lenses used in our analysis is shown as the black histogram in Fig.~\ref{fig:lens-redshift}, with a median redshift of approximately 0.2.

For systematics check, we also generate random catalogs separately for each lens area according to the established priority order. For surveys with similar depth (BOSS and SDSS-MGS), we use the random catalog from SDSS-MGS and reassign redshifts sampled from the lenses in that region. These individual random catalogs are then combined to create the random catalog for our fiducial lens sample. For each spectroscopic dataset, we use the corresponding random catalogs within the same sky regions, following the same survey prioritization applied to lens selection. These random catalogs are essential for estimating and correcting systematic effects in the weak lensing signal. Our lens sample is divided into redshift bins spanning the range \(0.1 < z < 0.5\), chosen to minimize redshift overlap with the HSC Y3 source galaxies. Table~\ref{tab:lenses} lists the redshift bins for the lenses, the number of galaxies in each bin, and the corresponding background galaxies from the source sample used in the analysis.

\subsubsection{SDSS}

The Sloan Digital Sky Survey (SDSS) I and II \cite{2000_York} covers an area of $11,663~\mathrm{deg}^2$, with imaging data collected using the 2.5-meter SDSS telescope. Photometry was performed in five broadband filters ($u$, $g$, $r$, $i$, $z$), and a subset of this imaging data was used to select targets for spectroscopic follow-up observations. The resulting spectroscopic galaxy sample includes galaxies with extinction and K+e-corrected Petrosian $r$-band magnitudes satisfying $r < 17.77$.

For our analysis, we utilize the New York University Value Added Galaxy Catalog (NYU VAGC), which compiles imaging as well as spectroscopic data from several publicly available surveys. These include SDSS \cite{2000_York,2008_Adelman}, the Two Micron All Sky Survey (2MASS) Point Source and Extended Source Catalogs \cite{2006_Skrutskie}, the IRAS Point Source Catalog Redshift Survey (PSCz) \cite{2000_Saunder}, the Two Degree Field Galaxy Redshift Survey (2dFGRS) \cite{2001_Colless}, the Third Reference Catalogue of Bright Galaxies (RC3) \cite{1991_De_vaucouleurs}, and the Faint Images of the Radio Sky at Twenty Centimeters survey (FIRST) \cite{1995_Becker}. The NYU VAGC spans $7,356~\mathrm{deg}^2$ with $r>14.5$, selecting a homogeneous sample of SDSS galaxies with reliable photometry and includes detailed treatments of survey geometry, $K+e$ corrections, peculiar velocities, and fiber collision effects, yielding a science-ready dataset for large-scale structure analyses. 

In this work, we specifically employ the \textit{safe} subset of the catalog\footnote{We use the \texttt{post\_catalog.dr72safe0} catalog from the NYU VAGC, available at \url{https://sdss.physics.nyu.edu/vagc/lss.html}}, which applies a uniform $r$-band flux limit of $14.5 < r < 17.6$ and includes nearest neighbour corrections for fiber collisions. For a comprehensive description of the catalog construction, we refer the reader to \citet{2005_Blanton}. We refer to this lens sample as the SDSS Main Galaxy Sample (SDSS MGS).

The NYU VAGC also provides a corresponding set of uniform random catalogs associated with the safe sample, which follow the survey footprint and account for the stellar mask. We construct random catalogs with a density 20 times that of the lens sample and assign redshifts by resampling (with replacement) from the redshift distribution of the SDSS MGS after the duplicates with overlapping deepper survey are removed. This sample spans a redshift range of $z < 0.3$, peaking at $z \sim 0.1$, as indicated by the red dashed histogram in Fig.~\ref{fig:lens-redshift}, which lies within the overlap with the HSC Y3 footprint.

In addition, we include lens galaxies at higher redshifts ($z < 0.55$), drawn from the large-scale structure spectroscopic catalog released as part of the SDSS-III \cite{2011_Eisenstein} Baryon Oscillation Spectroscopic Survey (BOSS) Data Release 12 \cite{2015_Alam}. These spectroscopic samples are selected based on SDSS photometry, extending the coverage of earlier SDSS-I/II imaging by targeting different populations for follow-up. BOSS adds approximately $3,000~\mathrm{deg}^2$ of imaging data which is subsequently used to also identify spectroscopic targets  in the southern sky. We utilize galaxies from the combined BOSS catalog, which includes the LOWZ sample ($0.15 < z < 0.35$) and the CMASS sample ($0.43 < z < 0.70$). To prevent duplication, we remove galaxies that are common to both the SDSS MGS and BOSS samples.

Due to the finite number of spectroscopic fibers, BOSS provides weights to correct for observational systematics. These include fiber collisions ($w_{\mathrm{cp}}$), redshift failures ($w_{\mathrm{noz}}$), and star-galaxy separation ($w_*$). The total weight $w_{\mathrm{l}}$ applied to each galaxy is given by $w_{\mathrm{l}} = w_* (w_{\mathrm{noz}} + w_{\mathrm{cp}} - 1)$ and for detailed description refer to \citet{2016_Reid}. We also use random catalogs with 20 times the number density of the BOSS lenses, selected from the associated BOSS randoms. The redshift distribution of the BOSS sample used in our analysis, restricted to the HSC Y3 footprint, is shown as the green dashed histogram in Fig.~\ref{fig:lens-redshift}.

\subsubsection{GAMA}
The Galaxy and Mass Assembly (GAMA) survey is a highly complete spectroscopic program conducted with the AAOmega spectrograph on the Anglo Australian Telescope. It provides spectra for approximately 300,000 galaxies down to a flux limit of $r < 19.8$, covering a sky area of $286~\mathrm{deg}^2$. In this work, we use data from GAMA Data Release 4 (DR4) \cite{2022_Driver}, specifically galaxies from the G3C catalogs version 10, which were used to construct the friends of friends group catalog in the GAMA equatorial regions and the G02 field \cite{2011_Robotham}. Among all the spectroscopic surveys included in our analysis, GAMA is the deepest, and as a result, the majority of our lens galaxies are selected from the GAMA fields (G02, G09, G12, and G15) that overlap with the HSC-Y3 footprint. We apply an additional redshift cut of $z < 0.55$ to define our lens sample to minimize the overlap with our primary target source redshift bin 3 and 4, which we are calibrating in our analysis. The redshift distribution of the selected GAMA galaxies is shown as the blue dashed histogram in Fig.~\ref{fig:lens-redshift}, with a median redshift of approximately $0.2$. To support the weak lensing analysis, we construct spatially uniform random catalogs with a density twenty times higher than that of the galaxy sample. These randoms cover the same sky area, and redshifts are assigned by sampling with replacement from the redshift distribution of the GAMA galaxies.

\subsubsection{DESI-BGS}

The Dark Energy Spectroscopic Instrument (DESI) is a sophisticated spectroscopic facility installed on the four-meter Mayall Telescope at Kitt Peak National Observatory \cite{2016_DESI_a}. It is equipped with robotic fiber positioners and has a field of view of approximately three degrees. DESI can simultaneously acquire spectra for more than 5,000 targets \cite{2016_DESI_b, 2023_Silber, 2024_Miller}, supported by imaging data from the publicly available DESI Legacy Imaging Surveys \cite{2017_Zou,2019_Dey}. It covers a wide wavelength range from the ultraviolet to the near-infrared. Over the course of its five-year survey, DESI aims to collect spectra for more than forty million galaxies across approximately 14,000 square degrees of the sky, probing redshifts up to \( z < 3.5 \).

In our fiducial analysis, we use galaxy data including sky position, redshift and associated weights from the DESI One Percent Survey \cite{2023_Hahn}, specifically the large-scale structure catalog released as part of DESI's early data release\cite{2024_DESI_a,2024_DESI_b}. We use the BGS Bright galaxy sample from this release, which is a uniform sample with a depth of \( r \leq 19.5 \), covering an area of 140 square degrees within the redshift range \( 0.05 < z < 0.5 \), and having a median redshift of 0.21\cite{2020_Ruiz}. The redshift distribution of these lenses is shown by the orange dashed histogram in Fig.~\ref{fig:lens-redshift}. We also use the corresponding set of random catalogs, with a density twenty times higher than the lens sample, provided by DESI.

In addition, we will compare our results by using the more recent Data Release 1 (DR1) from DESI~\cite{2025_DESI}, which includes all observations from the first thirteen months of the survey. This release also features uniformly reprocessed data that overlaps with the early data release. For consistency, we use the large-scale structure catalog from the BGS Bright sample and the associated random catalogs in the same manner as in our fiducial analysis. Since this data became publicly available after we unblinded our main results, we present the comparison of the inference from DESI-DR1 with the results from our fiducial analysis in Appendix~\ref{app:comp-DESI-DR1}. We refer readers to \citet{2023_Hahn} and \citet{2025_DESI} for a more comprehensive description of the DESI data.

\subsection{Sources: HSC-Y3 shape catalog and cosmic shear measurements}

The Hyper Suprime-Cam (HSC) is a wide-field optical camera located at the prime focus of the 8.2-meter Subaru Telescope on Maunakea \cite{2018_Miyazaki,2018_Komiyama, 2018_Furusawa,2018_Kawanomoto}. Its large field of view, combined with Subaru’s wide aperture and the excellent atmospheric condition at the site, makes it particularly well-suited for deep imaging surveys targeting weak gravitational lensing. The HSC Subaru Strategic Program (HSC-SSP) was allocated 330 nights to conduct a three-tiered imaging survey across multiple photometric bands, with the goal of balancing depth and area coverage \cite{2018_Aihara}.

The Wide layer of this survey, which was optimized for weak lensing cosmology, covers approximately 1,100 square degrees with multi-band imaging. Shape measurements of galaxies are carried out in the $i$-band, which achieves a $5\,\sigma$ depth of approximately $i\sim 26$. To ensure high-quality shape measurements, $i$-band exposures were preferentially taken under favorable seeing conditions, resulting in a median seeing of about 0.6 arc seconds.

For our analysis, we utilize the shape catalog derived from the S19a internal data release, which represents data processed at an intermediate stage between PDR2 and PDR3. This catalog was processed using \texttt{hscPipe} version 7 \cite{2022_Li_xianchong} and played a central role in the HSC-Y3 cosmological studies. Galaxies included in the catalog are drawn from the region with full-depth and full-color coverage in all five $grizy$ broad photometric bands. In addition to standard quality cuts based on pixel-level diagnostics, the shape catalog includes a series of selection criteria to ensure robust shape measurements. Specifically, extended sources are required to have extinction-corrected \(i\)-band \texttt{cmodel} magnitudes brighter than 24.5, an \(i\)-band signal-to-noise ratio (SNR) of at least 10, and a resolution factor exceeding 0.3. In addition a detection significance above \(5\sigma\) in at least two bands other than \(i\), and an aperture magnitude in the \(i\)-band below 25.5, measured within a 1 arcsecond diameter aperture, was also required to be part of the catalog. A comprehensive description of the catalog construction and selection methodology is provided in \citet{2022_Li_xianchong}.

The shape catalog covers a total area of 433 deg\(^2\), distributed across six survey fields: XMM, VVDS, GAMA09H, WIDE12H, GAMA15H, and HECTOMAP. It reaches an effective source galaxy number density of 19.9 arcmin\(^{-2}\), enabling high-precision weak lensing measurements. Shape calibration was performed using detailed image simulations \cite{2022_Li_xianchong}, generated with the \texttt{Galsim} package \cite{2015_Rowe}, and was supported by a comprehensive suite of tests to identify and control systematic biases affecting cosmological inference. Following the results of \citet{2023_Li_xianchong}, we exclude a region of approximately 20 deg\(^2\) in the GAMA09H field due to significant systematics identified in the HSC-Y3 cosmic shear analysis.

Galaxy shapes are measured using the re-Gaussianization point spread function (PSF) correction method \cite{2003_Hirata}, which estimates two ellipticity components \((e_1, e_2) = (e \cos 2\phi, e \sin 2\phi)\), where \(e = (1 - b^2/a^2)/(1 + b^2/a^2)\) is the distortion defined by the axis ratio \(b/a\), and \(\phi\) is the position angle of the major axis measured with respect to the equatorial coordinate system \cite{2002_Bernstein}. In addition to shape measurements, the catalog provides shear calibration parameters, including the multiplicative bias \(m\) and additive bias \(c\), necessary for accurate shear estimation.

HSC also provides photometric redshift information along with the shapes of individual galaxies, which have been computed using the dNNz photo-$z$ estimation method. The galaxies are divided into four redshift bins ranging from 0.3 to 1.5. These redshift bins have been jointly calibrated \cite{2023_Rau} using photometric redshift data and spatial cross-correlation with the CAMIRA Luminous Red Galaxy sample \cite{2014_Oguri,2018_Oguri,2021_Ishikawa}. The final two redshift bins have a limited overlap with the CAMIRA LRGs, and hence have to rely on purely the photometric information. For more details on the estimation of redshift distribution, we refer the reader to \citet{2023_Rau}. In this study, we select galaxies from each redshift bin for shear measurements. Instead of using the calibrated redshift distribution in our modeling, we adopt a blinding strategy based on shifts of these distributions, as discussed in Section \ref{sec:blinded sources}. 

We also incorporate real-space cosmic shear measurements and the corresponding covariance from \citet{2023_Li_xianchong}. These auto- and cross-correlation measurements and their covariances were calculated by using source galaxies divided into the same four redshift bins,  as we use in our analysis. These measurements, computed with the public software \texttt{TREECORR} \citep{2004_Jarvis}, cover an angular range of $7.1 < \theta/\mathrm{arcmin} < 56.6$ for $\xi_+$ and $31.2 < \theta/\mathrm{arcmin} < 248$ for $\xi_-$, and have a total signal-to-noise ratio of 26.6. The covariance matrices are constructed from 1404 mock shear catalogs based on the HSC survey, which account for variations in galaxy intrinsic shapes, image noise, and the cosmic shear signal \cite{2019_Shirasaki}. These mocks are generated using full-sky ray-tracing simulations assuming a WMAP9 cosmology.

We analyze the cosmic shear measurements using the blinded redshift distributions and compare the inferred results with those obtained from shear-ratio estimates. We also present constraints derived from the combined analysis.

\subsection{External dataset}
To break parameter degeneracies in our shear-ratio analysis and improve constraints on cosmological parameters—particularly the matter density and the source photometric redshift bias parameters $\Delta z$—we incorporate external baryon acoustic oscillation (BAO) measurements. These data help tighten constraints on the matter density and enable more precise estimation of the source photo-$z$ bias parameters.

We use BAO data from the final BOSS DR12 release~\citep{2017_Alam}, which is divided into three partially overlapping redshift slices centered at effective redshifts of 0.38, 0.51, and 0.61. These provide measurements of the angular diameter distance $D_M$ and the Hubble parameter $H$, obtained using the BAO method after applying density field reconstruction to mitigate non-linear degradation of the BAO signal.

The corresponding BOSS DR12 likelihood is implemented in \texttt{COSMOSIS}~\citep{2015_Zuntz}, which we use to jointly model the BAO data and our shear-ratio measurements as part of our fiducial cosmological analysis.

\section{Galaxy-galaxy lensing}\label{sec:meas}

We divide our lens sample into equally spaced redshift bins for both the \( z_{\mathrm{l}} < 0.4 \) and \( z_{\mathrm{l}} < 0.5 \) cases, as shown in Table~\ref{tab:lenses}. For the \( z_{\mathrm{l}} < 0.4 \) lenses, we use sources in redshift bins 2, 3, and 4 as background galaxies, and for the \( z_{\mathrm{l}} < 0.55 \) lenses, we use bins 3 and 4. The weak lensing signal is measured in five logarithmically spaced \(\theta\) bins, ranging from 0.17 to 137.5 arcminutes.

Weak gravitational lensing by foreground objects induces coherent distortions in the observed shapes of background galaxies \citep{2015_Kilbinger,2008_Hoekstra,2018_Mandelbaum}. The distortion $\gamma_{\rm t}$ is given by

\begin{equation}
\gamma_{\mathrm{t}} = \Sigma_{\mathrm{crit}}^{-1}(z_{\mathrm{l}}, z_{\mathrm{s}}) \Delta\Sigma, \label{eq:ggl}
\end{equation}

where $\Delta\Sigma = \bar{\Sigma}(<\theta) - \Sigma(\theta) \), with \( \bar{\Sigma}(<\theta) = \frac{\int_0^\theta \Sigma(\theta') 2\pi \theta' d\theta'}{\pi \theta^2} $ representing the average projected matter density enclosed within a disk of radius \( \theta \), and \( \Sigma(\theta) \) denoting the projected matter density. The term \( \Sigma_{\mathrm{crit}}^{-1}(z_{\mathrm{l}}, z_{\mathrm{s}}) \) is a geometrical factor that governs the lensing efficiency for a lens--source pair and is given by
\begin{equation}
\Sigma_{\mathrm{crit}}^{-1}(z_{\mathrm{l}}, z_{\mathrm{s}}) = \frac{4\pi G}{c^2} \frac{D_a(z_{\mathrm{l}}) D_a(z_{\mathrm{l}} , z_{\mathrm{s}}) (1 + z_l)^2}{D_a(z_{\mathrm{s}})},
\end{equation}
with \( \Sigma_{\mathrm{crit}}^{-1}(z_{\mathrm{l}}, z_{\mathrm{s}}) = 0 \) when \( z_{\mathrm{s}} < z_{\mathrm{l}} \), where \( z_{\mathrm{l}} \) and \( z_{\mathrm{s}} \) are the redshifts of the lens and source, respectively. Here, \( D_a(z_{\mathrm{s}}) \) is the angular diameter distance to the source galaxy, \( D_a(z_{\mathrm{l}}) \) is that to the lens galaxy, and \( D_a(z_{\mathrm{l}}, z_{\mathrm{s}}) \) is the angular diameter distance between the lens and the source and $G$ denotes the gravitational constant and $c$ is the speed of light.

Following \citet{2023_More}, we measure the stacked weak lensing signal \( \gamma_{\mathrm{t}} \) around galaxies at an angular separation \( \theta \) as
\begin{equation}
\gamma_{\mathrm{t}} = \frac{1}{1 + m} \left[ \frac{\sum_{\mathrm{ls} \in \theta} w_{\mathrm{ls}} e_{\mathrm{t, ls}}}{2\mathcal{R} \sum_{\mathrm{ls} \in \theta} w_{\mathrm{ls}}} - \frac{\sum_{\mathrm{ls} \in \theta} w_{\mathrm{ls}} c_{\mathrm{t, ls}}}{\sum_{\mathrm{ls} \in \theta} w_{\mathrm{ls}}} \right],
\end{equation}
where \( e_{\mathrm{t}} \) is the tangential component of the ellipticity, \( c_{\mathrm{t}} \) is the tangential component of the additive bias, and \( w_{\mathrm{ls}} = w_{\mathrm{l}} w_{\mathrm{s}} \) is the product of the lens weight \( w_{\mathrm{l}} \) and source weight \( w_{\mathrm{s}} \). The shear responsivity \( \mathcal{R} \) is given by
\begin{equation}
\mathcal{R} = 1 - \frac{\sum_{\mathrm{ls} \in \theta} w_{\mathrm{ls}} e^2_{\mathrm{RMS}}}{\sum_{\mathrm{ls} \in \theta} w_{\mathrm{ls}}},
\end{equation}
where \( e_{\mathrm{RMS}} \) is the RMS ellipticity  \cite{2002_Bernstein}. The multiplicative bias \( m \) is computed using the weighted average of the individual source multiplicative bias $m_{\rm s}$ and given as  
\begin{equation}
m = \frac{\sum_{\mathrm{ls} \in \theta} w_{\mathrm{ls}} m_{\mathrm{s}}}{\sum_{\mathrm{ls} \in \theta} w_{\mathrm{ls}}}.
\end{equation}
To account for selection effects near the edges of parameter cuts, we apply multiplicative and additive bias corrections following \citet{2022_Li_xianchong}:
\begin{equation}
\gamma_{\mathrm{t}} \rightarrow \frac{\gamma_{\mathrm{t}} - c_{\mathrm{sel}}}{1 + m_{\mathrm{sel}}},
\end{equation}
where
\begin{equation}
c_{\mathrm{sel}} = \frac{a_{\mathrm{sel}} \sum_{\mathrm{ls} \in \theta} w_{\mathrm{ls}} e_{\mathrm{t, psf}}}{\sum_{\mathrm{ls} \in \theta} w_{\mathrm{ls}}},
\end{equation}
and \( e_{\mathrm{t, psf}} \) is the tangential component of the PSF shape. The selection biases \( m_{\mathrm{sel}} \) and \( a_{\mathrm{sel}} \) depend on the fraction of galaxies at the selection threshold of aperture magnitude \( P(\mathrm{mag}_A) \) and resolution \( P(R_2) \), and are computed as
\begin{equation}
m_{\mathrm{sel}} = -0.05854 \, P(\mathrm{mag}_A = 25.5) + 0.01919 \, P(R_2 = 0.3),
\end{equation}
\begin{equation}
a_{\mathrm{sel}} = 0.00635 \, P(\mathrm{mag}_A = 25.5) + 0.00627 \, P(R_2 = 0.3).
\end{equation}
To eliminate residual scale-dependent systematics, we subtract the average tangential shear measured around random points. This correction is based on 20 realizations of random catalogs, matched in sky coverage and redshift distribution to our lens sample. We also verify that the cross-component of the shear signal is consistent with zero across all scales of interest, as shown in Appendix~\ref{app:x-shear}. Additionally, we examine the boost factors \citep{2005_Mandelbaum} to test for contamination from physically associated sources, finding them consistent with unity, indicating a clean background selection.

Since individual sources contribute to multiple $\theta$ bins, the resulting tangential shear measurements are inherently correlated. To estimate the shape-noise covariance, we use 300 realizations in which the source galaxy shapes are randomly rotated. Because the same set of lenses is used in all tangential shear measurements, the impact of large-scale sample variance largely cancels out, leaving shape noise as the primary source of uncertainty in our analysis\footnote{We have verified this using the BOSS LOWZ mock galaxy catalog and the associated HSC-Y3 shape catalog \cite{2019_Shirasaki}, finding percent level agreement between the jackknife-estimated covariance (which captures the contribution from large-scale structure) and the shape-noise covariance.}.. For consistency, we apply the same set of rotated realizations when computing the covariance for all lens redshift bins that share a common source redshift bin.

Shear measurements are performed around lenses in the redshift bins listed in Table~\ref{tab:lenses}, using sources in redshift bins 2, 3, and 4. Shear ratios are then computed by dividing the tangential shear from higher source redshift bins by that from lower ones. The covariance of the shear ratios is estimated by propagating the shape noise from the corresponding shear measurements, using the same set of randomly rotated catalogs to ensure consistency across all realizations.

Finally, we assess the potential cross-covariance between our shear ratio measurements and HSC-Y3 cosmic shear measurements. This is done using the same randomly rotated galaxies and the \texttt{TREECORR} \cite{2004_Jarvis} Python package for conducting the measurements, following the methodology described in \citet{2023_Li_xianchong}.

\subsection{Blinding}\label{sec:blinded sources}
The final cosmological analysis presented in this work is based on the methodology of \citet{2023_Li_xianchong}, who analyzed the HSC-Y3 cosmic shear measurements in real space. In that paper, blinding was carried out at the catalog level \citep[see also][]{2023_More} by constructing multiple catalogs with varying amount of multiplicative bias. Only one of these catalogs was correct. However, such a catalog level blinding would preserve shear ratios. Therefore, to avoid confirmation bias in our parameter inference, we adopted an alternative blinding strategy. 

Our analysis was blinded in model space by modifying the source redshift distributions. Specifically, we created four blinded versions by applying redshift shifts to the mean redshifts of the distributions provided by \citet{2023_Rau}. These shifts were designed to ensure that the resulting cosmological constraints remain consistent with the fiducial HSC-Y3 results, thereby preventing straightforward identification of the true distribution through comparison. To further guard against accidental unblinding, the shift values were encrypted using the public key of a team member who was not involved in the analysis. We applied blinding only to bins 3 and 4, since these were not robustly calibrated using the clustering redshift method. 

We carried out the full analysis independently for each blinded distribution, finalizing all results without access to the true redshifts. The analysis was unblinded only after the team reached consensus on the robustness and internal consistency of the results under all blinded configurations. We later confirmed that blinded version 1 corresponds to the true source redshift distribution, and the results presented in this work reflect the unblinded analysis.

\section{Model}\label{sec:model}
In this section, we describe our modeling framework for both shear-ratio and cosmic shear measurements. For the shear-ratio analysis, we follow the methodology of \citet{2018_Prat}, using the effective geometrical factor, which is sensitive to both the background cosmology and photometric redshift (photo-$z$) error parameters. For the cosmic shear analysis, we adopt the same modeling framework as used in the HSC Year 3 (HSC-Y3) analysis \cite{2023_Li_xianchong, 2023_Dalal}. This framework includes eleven physical parameters: five cosmological parameters, five intrinsic alignment parameters, and one baryonic feedback parameter. In addition, we include twelve systematic parameters: four photo-$z$ error parameters, four PSF systematics parameters, and four shear calibration bias parameters. The modeling choices are made specifically to enable a direct comparison with the HSC-Y3 results. The model parameters along with the priors are presented in the Table \ref{tab:parameters}.

\begin{table}
 \setlength{\tabcolsep}{20pt}
\begin{center}
\begin{tabular}{ll}  \hline\hline
{\bf Parameter} & {\bf Prior} \\ \hline
\multicolumn{2}{l}{\hspace{-1em}\bf Cosmological parameters
}\\\hline
$\Omega_{\mathrm{m}}$                 & $\mathcal{U}(0.1, 0.7)$\\
$A_{\mathrm{s}}\,(\times 10^{-9})$   & $\mathcal{U}(0.5, 10)$\\
$n_{\mathrm{s}}$                      & $\mathcal{U}(0.87, 1.07)$\\
$h_0$                                 & $\mathcal{U}(0.62, 0.80)$\\
$\omega_{\mathrm{b}}$                 & $\mathcal{U}(0.02, 0.025)$\\
\hline
\multicolumn{2}{l}{\hspace{-1em}\bf Baryonic feedback parameters
}\\\hline
$A_{\mathrm{b}}$                      & $\mathcal{U}(2, 3.13)$ \\\hline
\multicolumn{2}{l}{\hspace{-1em}\bf Intrinsic alignment parameters
}\\\hline
$A_1$                                 & $\mathcal{U}(-6, 6)$ \\
$\eta_1$                              & $\mathcal{U}(-6, 6)$ \\
$A_2$                                 & $\mathcal{U}(-6, 6)$ \\
$\eta_2$                              & $\mathcal{U}(-6, 6)$ \\
$b_{\mathrm{ta}}$                     & $\mathcal{U}(0, 2)$ \\
\hline
\multicolumn{2}{l}{\hspace{-1em}\bf Photo-$z$ systematics
}\\\hline
$\Delta z_1$                          & $\mathcal{N}(0, 0.024)$ \\
$\Delta z_2$                          & $\mathcal{N}(0, 0.022)$ \\
$\Delta z_3$                          & $\mathcal{U}(-1, 1)$ \\
$\Delta z_4$                          & $\mathcal{U}(-1, 1)$ \\\hline
\multicolumn{2}{l}{\hspace{-1em}\bf Shear calibration biases
}\\\hline
$\Delta m_1$                          & $\mathcal{N}(0.0, 0.01)$ \\
$\Delta m_2$                          & $\mathcal{N}(0.0, 0.01)$ \\
$\Delta m_3$                          & $\mathcal{N}(0.0, 0.01)$ \\
$\Delta m_4$                          & $\mathcal{N}(0.0, 0.01)$ \\\hline
\multicolumn{2}{l}{\hspace{-1em}\bf PSF systematics
}\\\hline
$\alpha'^{(2)}$                       & $\mathcal{N}(0, 1)$\\
$\beta'^{(2)}$                        & $\mathcal{N}(0, 1)$\\
$\alpha'^{(4)}$                       & $\mathcal{N}(0, 1)$\\
$\beta'^{(4)}$                        & $\mathcal{N}(0, 1)$\\\hline 
\multicolumn{2}{l}{\hspace{-1em}\bf Fixed parameters
}\\\hline
$\sum m_\nu\,(\mathrm{eV})$           & 0.06\\
$w$                                   & $-1$\\
$w_a$                                 & $0$\\
$\Omega_k$                            & $0$\\ 
$\tau$                                & $0.0851$\\\hline
\hline
\end{tabular}
\end{center}
\caption{Parameter priors used in the analysis. $\mathcal{U}(a,b)$ denotes a uniform distribution and $\mathcal{N}(\mu,\sigma)$ a normal distribution with mean $\mu$ and width $\sigma$.}
\label{tab:parameters}
\end{table}

\subsection{Shear ratios}\label{model:shear-ratios}
We calculate shear ratios between tangential shear measurements using lenses in redshift bin \(i\), with redshift distribution \(n_l^i(z_l)\), and source galaxies in redshift bin \(j\), with redshift distribution \(n_s^j(z_s)\). These distributions define an effective geometric factor as
\begin{equation}
\Sigma_{\mathrm{crit,eff}}^{-1\,i,j} = \iint \mathrm{d}z_l\, \mathrm{d}z_s\, n_l^i(z_l)\, n_s^j(z_s)\, \Sigma_{\mathrm{crit}}^{-1}(z_l, z_s).
\end{equation}
The shear ratio for lenses in redshift bin \(i\) and sources in two distinct redshift bins \(j\) and \(k\) is given by
\begin{equation}
\frac{\gamma_{t,j}}{\gamma_{t,k}} = \frac{\Sigma_{\mathrm{crit,eff}}^{-1\,i,j}}{\Sigma_{\mathrm{crit,eff}}^{-1\,i,k}},
\end{equation}
where \(\gamma_{t,j}\) is the tangential shear measured around lenses in bin \(i\) using sources in bin \(j\). Since both measurements share the same lens population, the excess surface density \(\Delta\Sigma\), which characterizes the matter distribution around the lenses, cancels in the ratio. This isolates the lensing efficiency factor \(\Sigma_{\mathrm{crit,eff}}^{-1\,i,j}\), which carries the cosmological dependence. As a result, the shear ratio is primarily sensitive to geometry and provides independent constraints on the photo-\(z\) error parameters.

\subsection{Cosmic shear}\label{sec:model-cosmic-shear}
We model the two-point correlation functions (2PCFs) of tomographic cosmic shear, $\xi_{\pm}^{ij}(\theta)$, for galaxy sources residing in redshift bins $i$ and $j$. This modeling follows the implementation by ~\citet{2023_Li_xianchong}, as provided in the publicly released package \texttt{COSMOSIS} \citep{2015_Zuntz}. Below, we summarize the essential components of the model and refer the reader to ~\citet{2023_Li_xianchong} for further details.

The cosmic shear 2PCFs can be expressed in terms of the $E$- and $B$-mode angular power spectra through a Hankel transform under the flat-sky approximation:
\begin{equation}
    \xi^{ij}_{\pm}(\theta) = \frac{1}{2\pi} \int d\ell \, J_{0/4}(\theta \ell) \left[ C^{E;ij}(\ell) \pm C^{B;ij}(\ell) \right],
\end{equation}
where $J_0$ and $J_4$ are the zeroth- and fourth-order Bessel functions of the first kind, respectively. These transforms are computed using the \texttt{FFTLog} \citep{2020_Fang} algorithm integrated within \texttt{COSMOSIS} \citep{2015_Zuntz}. The $E$-mode angular power spectrum receives contributions from both gravitational lensing and intrinsic alignments (IA) of galaxy shapes. These contributions can be decomposed as
\begin{equation}
    C^{E,ij} = C^{E,ij}_{\rm GG} + C^{E,ij}_{\rm II} + C^{E,ij}_{\rm GI} + C^{E,ji}_{\rm GI},
\end{equation}
where $C^{E,ij}_{\rm GG}$ denotes the auto-correlation due to lensing-induced shear, $C^{E,ij}_{\rm II}$ arises from intrinsic shape correlations, and $C^{E,ij}_{\rm GI}$ represents the cross-correlation between gravitational shear and intrinsic alignments. The intrinsic alignment terms originate from tidal interactions of galaxies with the surrounding large-scale structure. In line with ~\citet{2023_Li_xianchong}, we neglect the $B$-mode contributions, which are expected to be negligible given the scale cuts applied in the HSC-Y3 real-space analysis.

The lensing angular power spectrum, $C^{E,ij}_{\rm GG}$, is related to the nonlinear matter power spectrum, $P_{\rm m}(k,z)$, via the Limber approximation \cite{1953_Limber,2008_LoVerde} as follows:
\begin{equation}
    C^{E,ij}_{\rm GG} = \int_0^{\chi_H} d\chi \, \frac{q_i(\chi) q_j(\chi)}{\chi^2} \, P_{\rm m}\left(k = \frac{\ell + 1/2}{\chi}, z(\chi)\right),
\end{equation}
where $\chi$ is the comoving distance\footnote{In this work, we adopt a flat $\Lambda$CDM cosmology, so that the comoving angular diameter distance is equal to the comoving distance.}, $\chi_H$ is the horizon distance, and $q_i(\chi)$ is the lensing efficiency kernel for the $i$-th tomographic bin. This kernel is defined as
\begin{equation}
    q_i(\chi) = \frac{3}{2} \Omega_{\rm m} \left( \frac{H_0}{c} \right)^2 \frac{\chi}{a(\chi)} \int_\chi^{\chi_H} d\chi' \, n_i(\chi') \frac{\chi' - \chi}{\chi'},
\end{equation}
where $\Omega_{\rm m}$ and $H_0$ are the matter density parameter and the Hubble constant at redshift zero, respectively. The function $a(\chi)$ is the cosmological scale factor, and $n_i(z)$ represents the normalized redshift distribution of galaxies in the $i$-th redshift bin. The comoving distance $\chi$ is evaluated as a function of redshift, which allows computation of $P_{\rm m}(k,z)$ on a grid of redshift and wavenumber.

The linear matter power spectrum, $P_{\rm m}(k,z)$, is computed using the \texttt{BACCO}\citep{2021_Arico} emulator, trained on a suite of predictions generated by the Boltzmann solver \texttt{CLASS}\citep{2011_Lesgourgues,2011_Blas}. This solver computes the linear power spectrum based on cosmological parameters, including the total matter density $\Omega_{\rm m}$, the amplitude $A_s$ and spectral index $n_s$ of primordial curvature perturbations, the dimensionless Hubble parameter $h$, and the baryon density $\omega_b = \Omega_b h^2$. We adopt a fixed sum of neutrino masses, $\sum m_\nu = 0.06\,\mathrm{eV}$.


To incorporate nonlinear effects, we use the halo model-based \texttt{HMcode} \citep{2016_Mead}, calibrated against $N$-body and hydrodynamical simulations. Specifically, we adopt the 2016 version of \texttt{HMcode} to maintain consistency with the fiducial HSC Y3 cosmological analysis. This version introduces a halo bloating parameter $\eta_b$ and a scaling amplitude $A_b$, which modify the halo mass--concentration relation \citep{2015_Mead,2016_Mead} to model baryonic feedback \citep{2021_Asgari}. For computational efficiency, we fix $\eta_b$ using the empirical relation from \citet{2021_Joachimi}:
\begin{equation}
    \eta_b = 0.98 - 0.12 A_b.
\end{equation}
{ Although our choice of \texttt{HMcode-2016}\citep{2016_Mead} is motivated by consistency with fiducial choice for the HSC Y3 analyses, we note that an updated version, \texttt{HMcode-2020}\citep{2021_Mead}, with improved accuracy is also available and was tested in the HSC Y3 cosmology pipeline. The differences between the two versions have been found to have a negligible impact on the final results.}

%
This completes the specification of the matter power spectrum model used in our cosmic shear analysis. Using the Limber approximation, we similarly model the IA contributions, $C^{E,ij}_{\rm II}$ and $C^{E,ij}_{\rm GI}$, to the angular power spectrum for sources in redshift bins $i$ and $j$ as
\begin{align}
    C^{E,ij}_{\rm II}(\ell) &= \int_0^{\chi_H} d\chi \, \frac{n_i(\chi) n_j(\chi)}{\chi^2} \, P_{\rm II}\left(k = \frac{\ell + 1/2}{\chi}; z(\chi) \right), \\
    C^{E,ij}_{\rm GI}(\ell) &= \int_0^{\chi_H} d\chi \, \frac{q_i(\chi) n_j(\chi)}{\chi^2} \, P_{\rm GI}\left(k = \frac{\ell + 1/2}{\chi}; z(\chi) \right).
\end{align}
The $C^{E,ij}_{\rm II}$ and $C^{E,ij}_{\rm GI}$ terms describe the intrinsic–intrinsic and lensing–intrinsic shape correlations, respectively. These contributions are modeled using the tidal alignment and tidal torque (TATT) framework \cite{2019_Blazek}, which builds on nonlinear perturbation theory. Following the formalism of \citet{2022_Secco}, we adopt a version of the TATT model that includes only quadratic perturbation terms and introduces five key parameters: $A_1$, $A_2$, $\eta_1$, $\eta_2$, and $b_{\rm ta}$. Here, $A_1$ and $A_2$ determine the amplitude of the intrinsic alignment signal, with $A_1$ representing linear tidal alignment and $A_2$ capturing the quadratic tidal torquing contribution. The redshift dependence of these terms is parameterized by $\eta_1$ and $\eta_2$ with a pivot redshift of $z_0 = 0.62$, while $b_{\rm ta}$ accounts for the bias of the source galaxy population. This approach provides a flexible and physically motivated model of intrinsic alignments for weak lensing analyses.

\section{Systematics Modeling}
\label{sec:systematics}

In this section, we outline the modeling approach used to mitigate systematic uncertainties that could bias our cosmological inferences. Specifically, we account for potential biases arising from photometric redshift estimation, shear calibration, and residual point spread function (PSF) modeling errors.

Accurate modeling of the source redshift distribution in each of the four tomographic bins is essential for both shear-ratio and cosmic shear measurements. These redshift distributions are calibrated using spatial cross-correlations with luminous red galaxies (LRGs) in overlapping regions of the sky \citep[for more details refer to,][]{2023_Rau}. However, this calibration is only robust for bins 1 and 2 and partially for bin 3. No reliable calibration is available for bin 4. To account for uncertainties in the redshift distributions, we introduce a shift in the mean redshift of each bin parameterized by $\Delta z_i$ as
\begin{equation}
    n_i(z) \rightarrow n_i(z + \Delta z_i).
\end{equation}
We adopt informative Gaussian priors on $\Delta z_i$ for the first two bins, as recommended by ~\citet{2023_Rau}, and uninformative uniform priors for bins 3 and 4, following the fiducial HSC-Y3 analysis in ~\citet{2023_Li_xianchong}.

To model uncertainties in shear calibration, we introduce a multiplicative bias parameter $\Delta m_i$ for each redshift bin, which accounts for residual biases from the image simulation choices used to calibrate shear estimation. The two-point correlation functions (2PCFs) are corrected as
\begin{equation}
    \xi^{ij}(\theta) \rightarrow (1 + \Delta m_i)(1 + \Delta m_j) \, \xi^{ij}(\theta).
\end{equation}
We place Gaussian priors on $\Delta m_i$ with zero mean and a standard deviation of 0.01, in accordance with the calibration of the HSC-Y3 shear catalog \citep{2023_Li_xianchong}.

Finally, we account for PSF-related additive systematics arising from modeling errors in the shape measurement process. As demonstrated in mock studies by ~\citet{2023_Zhang}, inaccuracies in both the second- and fourth-order moments of the PSF and its residuals can introduce systematic biases. To correct for these effects, we include four parameters: $\alpha'^{(2)}$ and $\alpha'^{(4)}$ model additive biases due to second- and fourth-order PSF moments, while $\beta'^{(2)}$ and $\beta'^{(4)}$ capture the contribution from residual PSF-model moments. These corrections ensure that our shear measurements are robust against known PSF modeling uncertainties. We refer the reader to \citet{2023_Zhang} for more detailed discussion for the PSF systematics for the HSC-Y3 shape catalog analysis.

\section{Bayesian inference}\label{sec:bayesian}
We sample the posterior distribution for 23 model parameters, as listed in Table~\ref{tab:parameters}, which include cosmological, astrophysical, and systematic parameters. We denote the set of parameters by $\Theta$, and given the data vector $\mathcal{D}$, the posterior $\mathcal{P}(\Theta|\mathcal{D})$ is expressed using Bayes theorem as
\begin{equation}
    \mathcal{P}(\Theta|\mathcal{D}) \propto \mathcal{P}(\mathcal{D}|\Theta)\, \mathcal{P}(\Theta),
\end{equation}
where $\mathcal{P}(\mathcal{D}|\Theta)$ is the likelihood, and $\mathcal{P}(\Theta)$ represents the prior distributions, as specified in Table~\ref{tab:parameters}. 
 Following \citet{2022_Sanchez}, we adopt a Gaussian likelihood with a covariance matrix corrected by the Hartlap factor \citep{2007_Hartlap} to obtain an unbiased estimate of the inverse covariance matrix in the presence of noise.

For Bayesian inference, we employ the nested sampler \texttt{PolyChord} \citep{2015_Handley}, following the approach used in \citet{2023_Li_xianchong} for the fiducial analysis. In our setup, we use 500 live points, 20 number of repeats, and a sampling efficiency tolerance of one percent. In this work, we report the median of the posterior distribution for each parameter, along with the 68 percent confidence interval around it.

We independently analyze our shear ratios and cosmic shear measurements, and compare the resulting parameter constraints to assess consistency between these two independent probes. The shear ratios are sensitive to the background cosmological parameters and source redshift shift parameters, $\Delta z$, as described in Sec.~\ref{model:shear-ratios}, and thus provide an independent constraint on these parameters. Furthermore, we also derive parameter constraints from the combined analysis of shear ratios and cosmic shear.


\section{Tangential shear profiles}\label{sec:shear}
\begin{figure*}
    \centering
    \includegraphics[width=\textwidth]{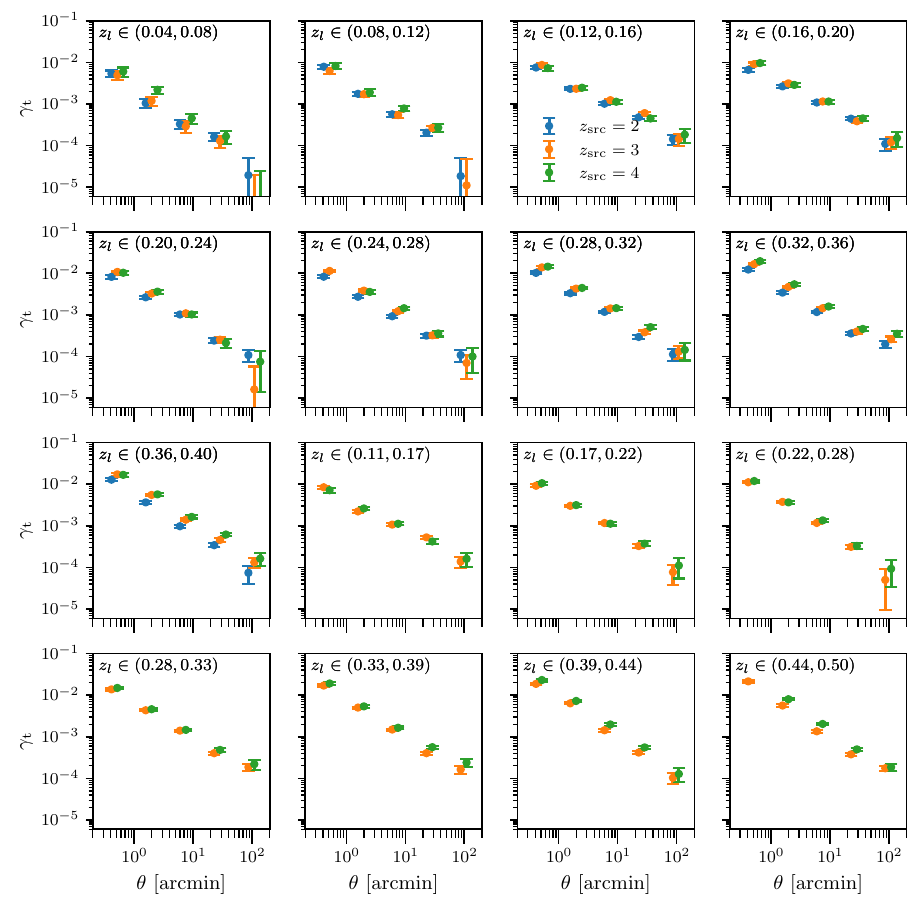}
    \caption{\textit{Tangential shear measurements:} Shear signals with error bars estimated from 300 random rotations. Blue, orange, and green points correspond to source redshift bins two, three, and four, respectively. A small horizontal offset in $\theta$ facilitates comparison, and each panel indicates the lens redshift bin.}
    \label{fig:shear-meas}
\end{figure*}

We divided our lens sample into thin redshift bins, as shown in Table~\ref{tab:lenses}, and measured the weak lensing signal around them using the shapes of background galaxies selected from the tomographic source redshift bins employed in the HSC-Y3 fiducial cosmological analysis. The tangential shear signal, measured across five logarithmically spaced angular separation bins ranging from $0.2$ to $200$ arcminutes, is presented in Fig.~\ref{fig:shear-meas}. To account for residual PSF systematics, we subtracted the signal measured around a set of random points with 20 times more entries than the lens sample \citep[][]{2004_Sheldon, 2005_Mandelbaum, 2017_Singh}.

Each panel shows the shear signal with data points representing our measurements and error bars estimated from shape noise using 300 random rotations. The blue, orange, and green points correspond to shear profiles measured using background galaxies in source redshift bins two, three, and four, respectively. To aid comparison across redshift bins, we applied a small horizontal offset in angular separation $\theta$.

\begin{figure*}
    \centering
    \includegraphics[width=\textwidth]{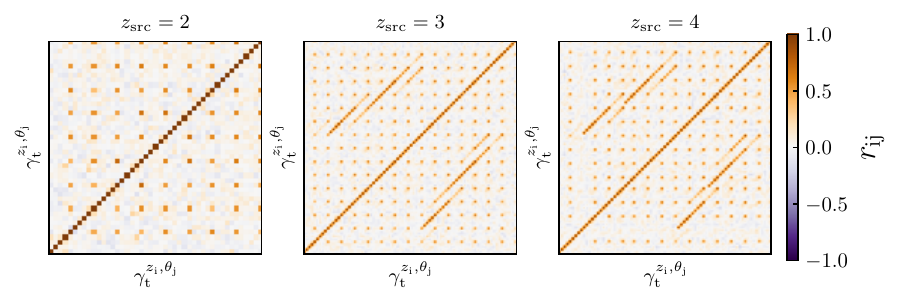}
\caption{\textit{Shear covariance:} Normalized covariance matrices of the tangential shear measurements for each source redshift bin. The quantity $\gamma_{\rm t}^{z_i,\theta_j}$ represents the tangential shear measured in the $i$-th lens redshift bin (see Table~\ref{tab:lenses}) at angular bin $\theta_j$, with correlation coefficient $r_{ij}$.}

    \label{fig:shear-covariance}
\end{figure*}

The shear signal declines with increasing angular separation, reflecting the weakening distortion on background galaxies at larger distances from the lens. The variation in amplitude across different source redshift bins for each lens bin highlights the redshift dependence of the background galaxies, driven by the geometric lensing efficiency factor, $\Sigma_{\rm crit}^{-1}$, defined in Eq.~\ref{eq:ggl}.

Fig.~\ref{fig:shear-covariance} shows the normalized covariance matrices for the shear measurements, computed for sources in the different redshift bins. The notation $\gamma_{\rm t}^{z_i, \theta_j}$ refers to the shear signal measured in the $i^{\rm th}$ lens redshift bin (see Table~\ref{tab:lenses}) at angular bin $\theta_j$, with correlation coefficient $r_{\rm ij}$. Shear measurements using sources in redshift bin two are largely uncorrelated, with only mild correlations at larger angular scales. For sources in redshift bins three and four, we observe mild correlations due to shared lenses across overlapping redshift bins.

\begin{figure}
    \centering
    \includegraphics[width=\columnwidth]{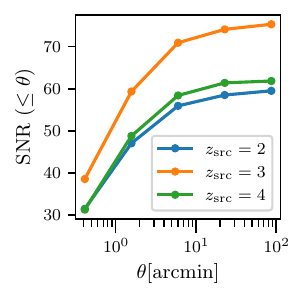}
\caption{\textit{Cumulative signal-to-noise ratio:} Solid blue, orange, and green lines show the cumulative signal-to-noise ratio for the tangential shear profiles as a function of angular separation $\theta$ for source redshift $z_{\rm src}$ bins two, three, and four, respectively.}

    \label{fig:shear-cum-snr}
\end{figure}

In Fig.~\ref{fig:shear-cum-snr}, we present the cumulative signal-to-noise ratio (SNR) for each source redshift bin, based on the angular scale cuts used in our analysis. We obtain SNR values of approximately $59$, $75$, and $62$ for sources in redshift bins two, three, and four, respectively. These differences reflect the variation in shear sensitivity with the redshifts of the sources and lenses, which have a median redshift near $0.2$, and are influenced by the lensing efficiency factor, $\Sigma_{\rm crit}^{-1}$.

{To assess potential systematics, we examined the cross-component of the shear profiles around the lenses. As shown in Fig.~\ref{fig:shear-cross-meas} of Appendix~\ref{app:x-shear}, we detect no statistically significant signal, with the $p$ value remaining above our adopted threshold of 2 percent after subtracting the signal measured around random points. We also examine the boost factors \citep{2005_Mandelbaum} to test for contamination from physically associated sources, finding them consistent with unity, which indicates a clean selection of background galaxies.}

 The magnification bias can also affect the measured shear induced by lens galaxies on background sources. In our analysis, the median redshift for galaxies in any source bin differs by at least 0.35 from the redshift of galaxies in the lens redshift bins. For such a difference shear ratios can be affected 
sub-percent level which is well within the statistical uncertainties of our measurements \citep{2019_Unruh}. Nonetheless, a proper treatment of magnification bias requires dedicated mock analyses and is warranted in future work.

\section{Results}

\subsection{Shear ratios}\label{sec:shear-ratio}
\begin{figure*}
    \centering
    \includegraphics[width=\textwidth]{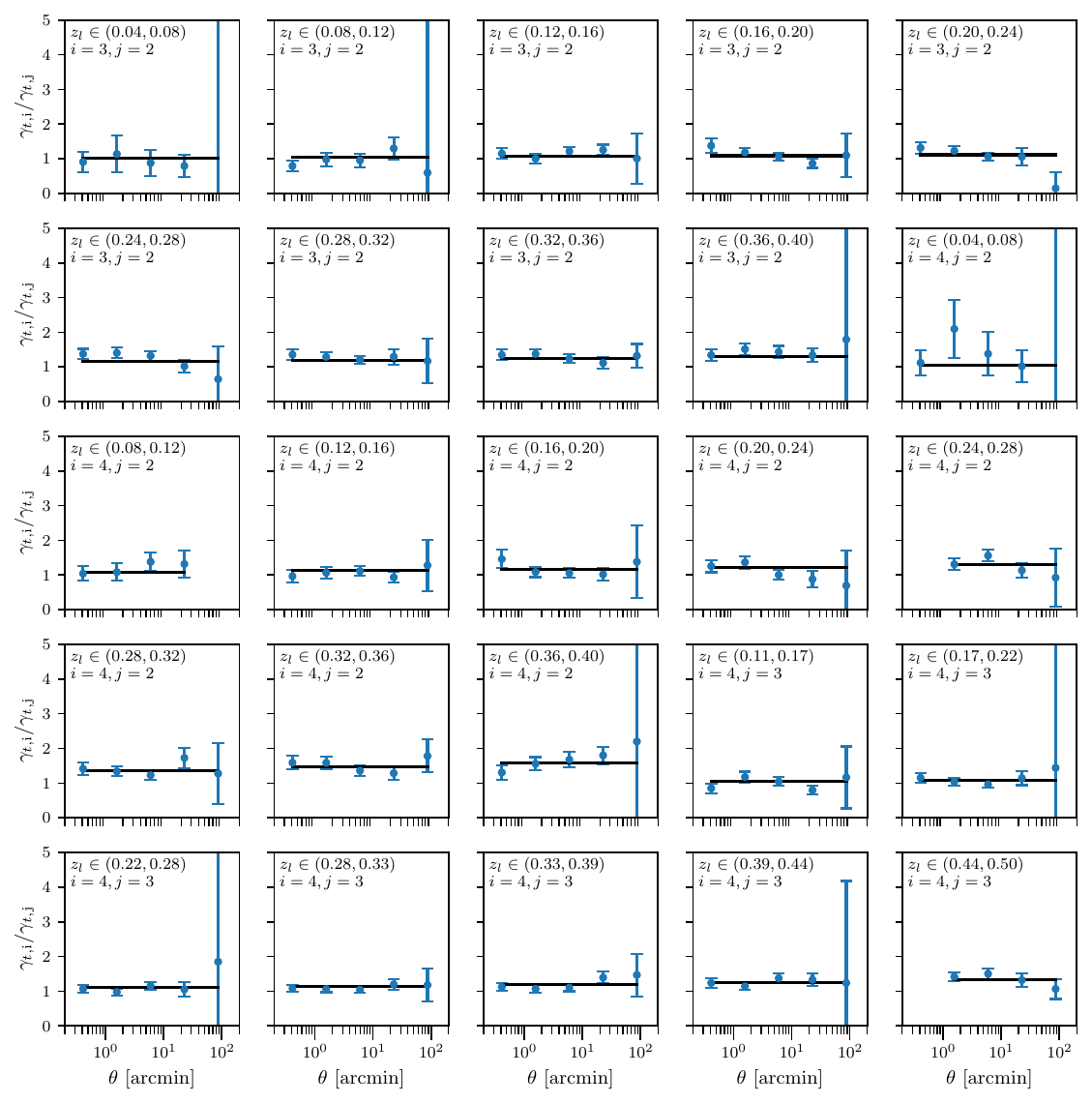}
\caption{\textit{Shear ratio measurements:} Blue points with shape noise error bars show the measured shear ratios for each lens bin. The black line indicates the best fit prediction from the fiducial model. Indices $i$ and $j$ on the $y$-axis denote the source redshift bins used in the numerator and denominator of the shear ratio, respectively.}
    \label{fig:shear-ratio-meas}
\end{figure*}

We compute the ratio of tangential shear profiles shown in Fig.~\ref{fig:shear-meas}, measured around lens galaxies in different redshift bins using background sources located in separate redshift bins. The resulting shear ratio measurements are shown in Fig.~\ref{fig:shear-ratio-meas} as blue data points with error bars corresponding to each lens redshift bin. The solid black line represents the best-fit prediction obtained through nested sampling of the Bayesian inference using a Gaussian likelihood, as described in Sec.~\ref{sec:bayesian}. This prediction is based on our fiducial model, which incorporates the effective lensing efficiency geometrical factor $\Sigma^{-1}_{\mathrm{crit\,eff}}$, detailed in Sec.~\ref{model:shear-ratios}.

The error bars represent the contribution from shape noise and are estimated from 300 realizations in which the shapes of source galaxies are randomly rotated. This randomization removes the tangential shear signal induced by the lenses. However, directly computing shear ratios from these randomized realizations can introduce biases in the covariance estimation due to the presence of small residual signals in the denominator. To mitigate such biases, we add the randomized shear signal to the true tangential shear measurements and treat the resulting profiles as individual realizations.

\begin{figure}
    \centering
    \includegraphics[width=\columnwidth]{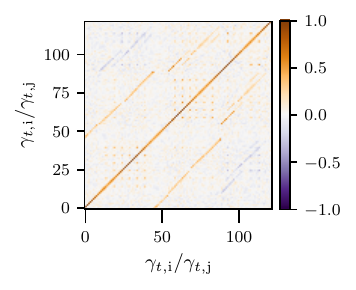}
    \caption{\textit{Shear ratio covariance:} Normalized covariance matrix for the shear ratio measurements shown in Fig.~\ref{fig:shear-ratio-meas}.}
    \label{fig:shear-ratio-cov}
\end{figure}

The corresponding normalized covariance matrix is presented in Fig.~\ref{fig:shear-ratio-cov}. We observe mild correlations between different lens redshift bins, which originate from correlations in their tangential shear measurements, as shown in Fig.~\ref{fig:shear-covariance}. After applying the Hartlap correction~\cite{2007_Hartlap}, we obtain a total signal-to-noise ratio of 203.

We have also checked the covariance between the shear ratio measurements and the cosmic shear measurements, using the same set of randomly rotated galaxy shapes. We find no significant correlation between the two observables, indicating that the shear ratio and cosmic shear measurements can be treated independently and combined in a joint analysis.

\subsection{Fiducial constraints}\label{sec:fiducial-constraints}

In this section, we present the fiducial parameter constraints from our shear-ratio measurements and compare them with those obtained using the HSC-Y3 cosmic shear measurements. We derive the constraints by combining our shear-ratio measurements with the BOSS DR12 BAO data as our fiducial setup. We first present the constraints from the independent analyses and subsequently discuss the cosmological implications from the joint analysis.

Our shear-ratio measurements are modeled using the methodology described in Sec.~\ref{sec:model}. The solid black line in Fig.~\ref{fig:shear-ratio-meas} shows the best-fit model, which yields a minimum $\chi^2$ of 84.30 with an effective number of degrees of freedom of 124.4 and a $p$-value of 99 percent. The effective degrees of freedom are computed following Eq.~(29) from \citet{2019_Raveri}. For consistency, we also apply the modeling framework described in Sec.~\ref{sec:model-cosmic-shear} to the HSC-Y3 cosmic shear measurements, obtaining a best-fit $\chi^2$ of 146.5 with 132.5 effective degrees of freedom and a $p$-value of 19 percent.

{As an independent consistency check, we fit the shear-ratio measurements with a simpler constant model to test for potential radial dependence. This test is performed independently of the fiducial modeling approach and shows no evidence for significant scale dependence, with $p$-values exceeding 30 percent. All results and conclusions presented in this work are therefore based exclusively on the fiducial shear-ratio model.}

{In addition, the shear-ratio measurements could also be used to constrain $dz_2$ separately. However, allowing $dz_2$ to vary introduces additional degeneracies with $dz_3$ and $dz_4$, substantially weakening the resulting constraints. Our aim is to calibrate the redshift estimates which rely solely on photometry. We therefore focus our analysis on the third and fourth source bins, which are known to have difficulties in calibration using cross-correlation techniques \citep{2023_Rau}. Thus, they benefit most from the inclusion of shear-ratio information. For consistency and to enable a direct comparison, we adopt the same priors as in the fiducial analysis of \citet{2023_Li_xianchong}, and report constraints on the redshift parameters of source bins~3 and~4.}

\begin{figure}
    \centering    \includegraphics[width=\columnwidth]{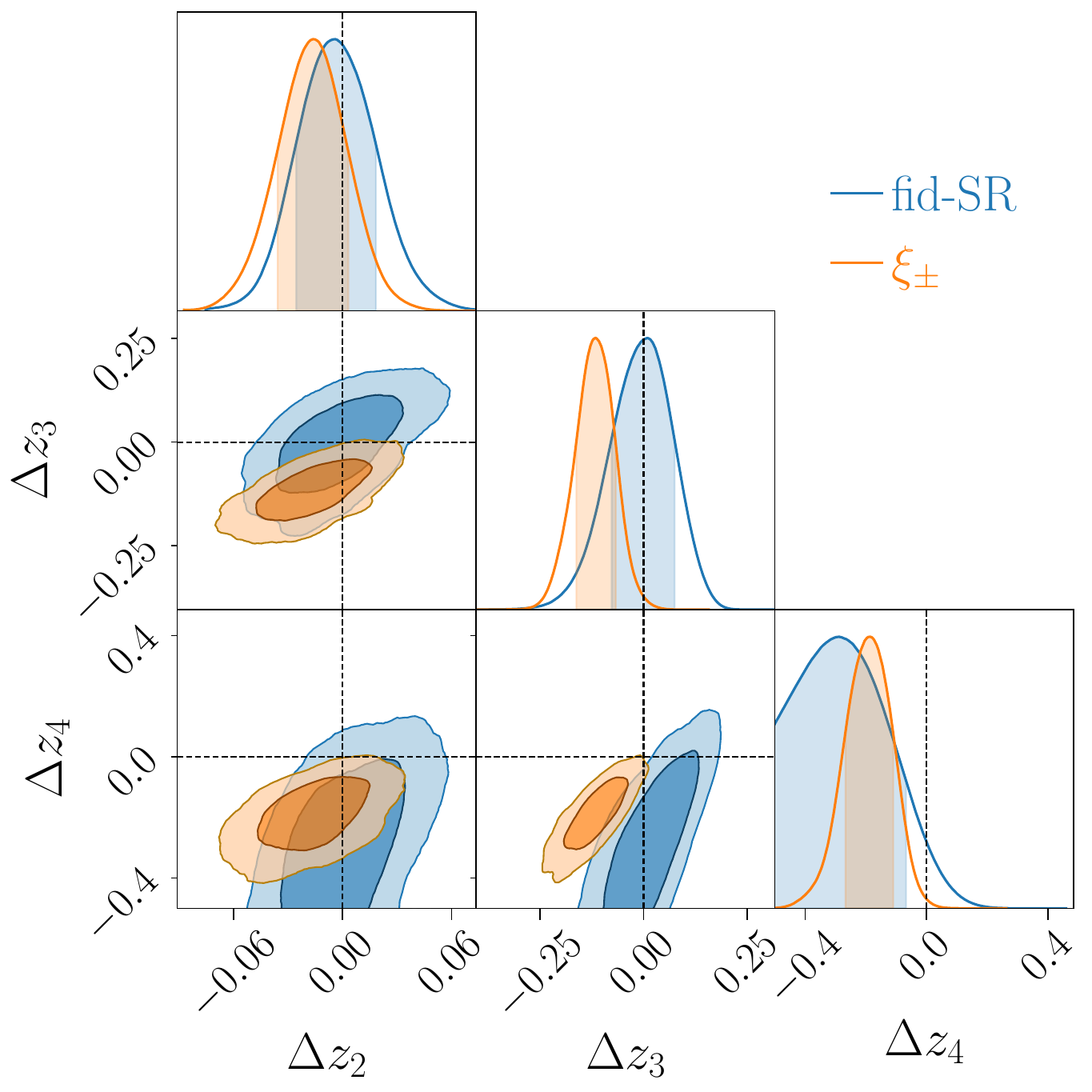}
    \caption{\textit{Photo-$z$ biases:} Blue contours show constraints from our fiducial shear-ratio analysis, while orange contours correspond to constraints inferred from modeling the HSC-Y3 real-space cosmic shear measurements. The contours denote the 68\% and 95\% confidence intervals. The black dashed vertical and horizontal lines indicate the biases equals to zero.}

    \label{fig:photo-z-biases}
\end{figure}

In Fig.~\ref{fig:photo-z-biases}, we present constraints on the photometric redshift bias parameters $\Delta z_2$, $\Delta z_3$, and $\Delta z_4$ for source redshift bins 2, 3, and 4, respectively. The contours represent the 68\% and 95\% confidence intervals of the posterior distributions from our analyses. Blue contours correspond to the fiducial shear-ratio analysis combined with BOSS BAO measurements, while orange contours show results from the independent HSC-Y3 cosmic shear analysis. These parameter constraints are summarized in Table~\ref{tab:fiducial-constraints}. We further compared the constraints on the remaining parameters and found them to be consistent with the expectations from the cosmic shear analysis.


{The fiducial shear-ratio analysis provides constraints on the photometric redshift biases that are consistent with those obtained from cosmic shear measurements.}
The posterior distributions overlap within approximately $1.2\sigma$, although a modest difference in the joint $\Delta z_3$–$\Delta z_4$ contours suggests potential areas for further study. These results demonstrate that the fiducial shear-ratio analysis provides independent and complementary constraints on source photometric redshift biases relative to cosmic shear.  The weaker constraint on $\Delta z_4$ arises from the lower signal-to-noise ratio measurements using background galaxies residing in the source bin four as shown in Fig. \ref{fig:shear-cum-snr}. {Furthermore, \citet{2023_Rau} derive informative Gaussian priors for the source redshift uncertainties in bins~3 and~4, given by $\mathcal{N}(0,\,0.031)$ and $\mathcal{N}(0,\,0.034)$, respectively. Our constraints are in agreement with those of \citet{2023_Rau} and are fully consistent within the quoted uncertainties.}

A corresponding analysis using DESI DR1 galaxies is presented in Appendix~\ref{app:comp-DESI-DR1} and shown in Fig.~\ref{Appendix:fig:photo-z-biases}. In this case, the $\Delta z_3$–$\Delta z_4$ contours do not exhibit the same tilt observed in the fiducial analysis. This difference may arise from the heterogeneous nature of the fiducial lens sample, which combines galaxies from multiple surveys of varying depths. In contrast, the DESI DR1 galaxies provide more uniform and contiguous coverage within the HSC-Y3 footprint. We speculate that the dominant lensing contribution from GAMA is responsible for the tilt seen in the fiducial contours.

{Further investigation indicates that HSC source galaxies in regions overlapping with the GAMA footprint, which covers an area of about $132\,\mathrm{deg}^2$, tend to prefer slightly higher values of $\Delta z_3$. To test this, we excluded lens galaxies from DESI DR1 residing in the GAMA fields and repeated the analysis. This exclusion resulted in a lower best-fit value of $\Delta z_3$, bringing it into closer agreement with the value inferred from the HSC-Y3 cosmic shear analysis. These results suggest that discrepancies arising from heterogeneous lens samples can average out when using a more uniform sample, such as DESI DR1, which has a larger overlapping area of $390\,\mathrm{deg}^2$. This test indicates such variations may exist, but a more detailed investigation is warranted in future work.}

To assess potential tension between the $\Delta z$ constraints from the HSC-Y3 cosmic shear and our fiducial shear-ratio analysis, we used the \texttt{tensiometer} package \citep{2021_Raveri}. The results indicate agreement at the $1\sigma$ level. Encouraged by this consistency, we proceed with a combined analysis to derive constraints on cosmological parameters. The joint model yields a best-fit $\chi^2$ of 241.2 with an effective number of degrees of freedom of 258.81 and {a $p$-value of 78 percent}, indicating a good fit.

\begin{figure}
    \centering
    \includegraphics[width=\columnwidth]{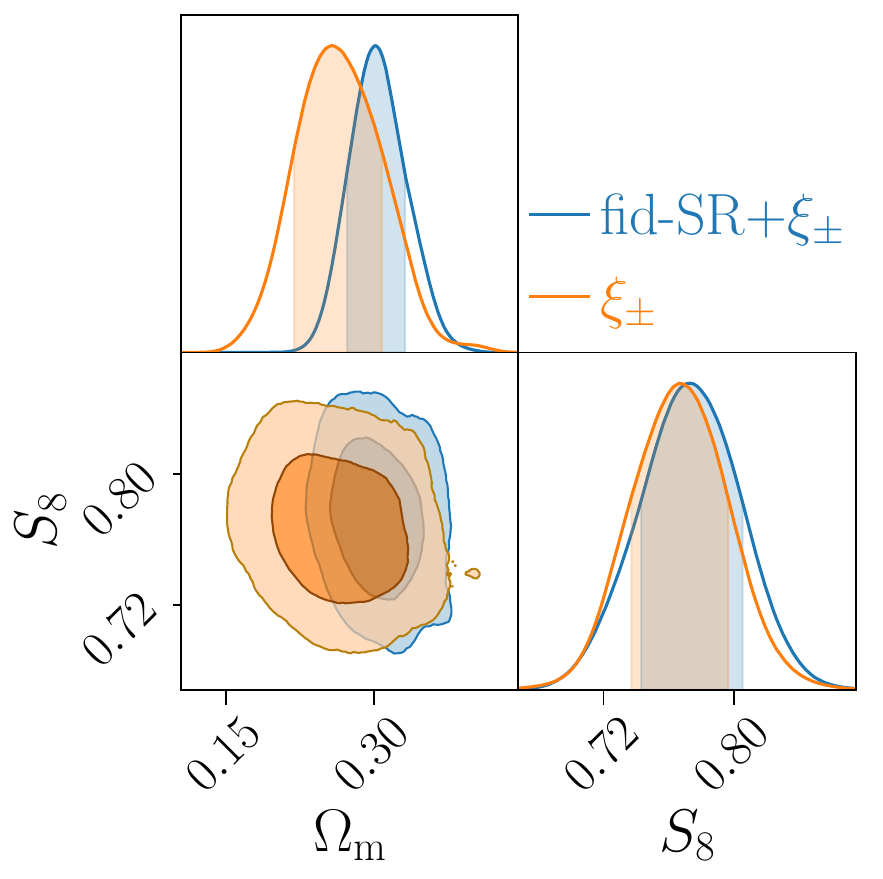}
    \caption{{\it Cosmological constraints:} Comparison of cosmological constraints from the combined analysis of HSC-Y3 cosmic shear and shear-ratio measurements. The blue contours show the results obtained using the fiducial shear-ratio data combined with cosmic shear, while the orange contours represent the constraints derived from modeling the HSC-Y3 cosmic shear measurements alone.The contours denote the 68\% and 95\% confidence intervals.}

    \label{fig:cosmo-constraints}
\end{figure}

The resulting constraints on the cosmological parameters $\Omega_{\rm m}$ and $S_8$ are reported in Table~\ref{tab:fiducial-constraints}. Figure~\ref{fig:cosmo-constraints} presents the one-dimensional posterior distributions and two-dimensional marginalized contours. These results demonstrate consistency between the HSC-Y3 cosmic shear only analysis (orange contours) and the combined analysis using our fiducial shear-ratio measurements and HSC-Y3 cosmic shear data (blue contours). The slight difference in the $\Omega_{\rm m}$ parameter arises from the inclusion of the BAO measurements, and these differences are consistent with those found in the HSC-Y3 analysis \citep{2023_Li_xianchong} after the inclusion of the BAO dataset. 
Additional comparisons using DESI DR1 lens galaxies are shown in Appendix~\ref{app:comp-DESI-DR1}, with Fig.~\ref{Appendix:fig:cosmology-constraints} further demonstrating the agreement.

\begin{table}
\setlength{\tabcolsep}{3pt}
\begin{center}
\begin{tabular}{ccc}  
\hline
{\bf Photo-$z$ bias} & {\bf fid-SR}  & {\bf $\xi_{\pm}$}\\
\hline
$\Delta z_2$ & $-0.004_{-0.02}^{+0.022}$ & $-0.016_{-0.02}^{+0.018}$\\
$\Delta z_3$ & $-0.002_{-0.082}^{+0.071}$ & $-0.117_{-0.048}^{+0.046}$\\
$\Delta z_4$ & $-0.34_{-0.263}^{+0.198}$ & $-0.19_{-0.081}^{+0.077}$\\
\hline
\hline
{\bf Cosmology } & {\bf fid-SR + $\xi_{\pm}$}  & {\bf $\xi_{\pm}$}\\
\hline
$\Omega_{\rm m}$   & $0.304_{-0.029}^{+0.03}$ & $0.264_{-0.042}^{+0.047}$\\
$S_8$   & $0.773_{-0.031}^{+0.031}$
 & $0.767_{-0.03}^{+0.029}$\\ 
\hline
\end{tabular}
\end{center}
\caption{\textit{Parameter constraints:} Constraints on photo-$z$ bias parameters ($\Delta z_{\rm i}$) and cosmological parameters ($\Omega_{\rm m}$ and $S_8$) from the fiducial and combined fiducial shear-ratio and cosmic shear analyses.}
\label{tab:fiducial-constraints}
\end{table}

\section{Summary and outlook}\label{sec:summary}
In this work, we use small-scale lensing measurements to place independent constraints on the photometric redshift (photo-\(z\)) calibration parameters for source galaxies. Specifically, we analyze galaxies from the HSC-Y3 survey spanning roughly $300 \, {\rm deg}^2$, divided into four tomographic bins, and measure the tangential shear around foreground lenses selected from several spectroscopic large-scale structure surveys. These measurements are then used to compute shear ratios for a finely binned lens redshift sample.

To avoid confirmation bias, we adopt a blinded analysis strategy in which small shifts are applied to the input source redshift distributions. All constraints were derived in the blinded phase, results were frozen, and the analysis was only unblinded afterward. Because the shear ratio is primarily sensitive to background cosmology and photo-\(z\) systematics, it provides an effective means of constraining the mean redshift shift parameters for source bins 2, 3, and 4.

We then compare our results with constraints obtained from modeling the HSC-Y3 real-space cosmic shear measurements. A summary of our findings is as follows:

\begin{itemize}
    \item \textbf{Lens Sample Construction}: We constructed the lens sample using spectroscopic galaxies from GAMA, SDSS-MGS, SDSS-BOSS, and DESI-EDR, all within the HSC-Y3 footprint and with redshifts below 0.5. The lenses were divided into fifteen narrow redshift bins. Associated random catalogs were included to ensure accurate statistical analysis.

    \item \textbf{Tangential Shear Measurements}: We measured the tangential shear profile around the lens galaxies in five logarithmically spaced angular bins ranging from 0.2 to 200 arcminutes. Background source galaxies were drawn from four HSC-Y3 tomographic bins. For source bins 2, 3, and 4, we achieved signal-to-noise ratios of 59, 75, and 62, respectively.

    \item \textbf{Shear Ratio Computation}: These shear measurements were used to compute shear ratios across the lens bins, yielding a combined signal-to-noise of 203. This shear-ratio dataset, when combined with BAO measurements from BOSS-DR12, constitutes the fiducial data vector for our analysis.

    \item \textbf{Photometric Redshift Bias Constraints}: We obtained independent constraints on the photometric redshift bias parameters \(\Delta z\), focusing on source bins 2, 3 and 4, which are only partially calibrated using clustering redshifts and thus more susceptible to photo-\(z\) systematics. We found
    \(\Delta z_2 = -0.004_{-0.02}^{+0.022}\), \(\Delta z_3 = -0.002_{-0.082}^{+0.071}\) and \(\Delta z_4 = -0.34_{-0.263}^{+0.198}\), consistent with constraints from HSC-Y3 cosmic shear modeling: \(\Delta z_2 = -0.016_{-0.02}^{+0.018}\), \(\Delta z_3 = -0.117_{-0.048}^{+0.046}\) and \(\Delta z_4 = -0.19_{-0.081}^{+0.077}\).

    \item \textbf{Systematic Test with Homogeneous Lenses}: We observed a mild shift in the confidence contours in the \(\Delta z_3\)--\(\Delta z_4\) plane, likely due to the heterogeneous depth of the spectroscopic lens sample. To test this, we repeated the analysis using only DESI-DR1 lenses, which provide more uniform coverage across the HSC-Y3 footprint. In this case, the contour tilt was no longer present.

    \item \textbf{Joint Cosmological Constraints}: Finally, we combined our fiducial shear-ratio measurements with HSC-Y3 cosmic shear data to derive joint constraints on cosmological parameters. We obtained \(\Omega_{\rm m} = 0.304_{-0.029}^{+0.03}\) and \(S_8 = 0.773_{-0.031}^{+0.031}\), which are consistent within \(1\sigma\) with the cosmic shear-only results: \(\Omega_{\rm m} = 0.264_{-0.042}^{+0.047}\) and \(S_8 = 0.767_{-0.03}^{+0.029}\).
\end{itemize}

Our analysis highlights the potential of the shear-ratio test as a powerful probe for constraining biases in the source redshift distribution, thereby improving the precision of cosmological parameter estimates. Since shear ratios are primarily sensitive to geometry, they offer significant advantages for investigating evolving dark energy models, as recently suggested by galaxy clustering results from the DESI survey \citep{2025_Adame,2025_Desi_karim}, {and can be beneficial for the upcoming HSC final-year analysis}. Furthermore, future surveys such as LSST \citep{2019_Ivezi}, Roman \citep{2015_Spergel}, and Euclid \citep{2011_Laureijs} will greatly enhance the impact of shear-ratio analyses, providing more stringent tests of the $\Lambda$CDM paradigm and helping to refine its key parameters, which currently exhibit persistent tensions with early-universe Planck-CMB observations.

\begin{acknowledgments}
We acknowledge the use of the high performance com-
puting facility - Pegasus at IUCAA. DR acknowledges funding from the European Research Council (ERC) under the European Union’s Horizon 2020 research and innovation program (Grant agreement No. 101053992). DR also would like to thank the University Grants Commission (UGC), India, for financial support as a senior research fellow during the initial phase of the work.

This work was supported in part by World Premier International Research Center Initiative (WPI Initiative), MEXT,
Japan. HM is supported by JSPS KAKENHI Grant Numbers J20H01932, J22K21349, and J23H00108. TZ is supported by Schmidt Sciences. SS is supported by the JSPS Overseas Research Fellowships. MS is supported by JSPS KAKENHI Grant Numbers JP24H00215 and JP24H00221. Numerical computations were [in part] carried out on the analysis servers and the general-purpose PC cluster at
the Center for Computational Astrophysics, National Astronomical Observatory of Japan.

The Hyper Suprime-Cam (HSC) collaboration includes the astronomical communities of Japan and Taiwan, and Princeton University. The HSC instrumentation and software were developed by the National Astronomical Observatory of Japan (NAOJ), the Kavli Institute for the Physics and Mathematics of the Universe (Kavli IPMU), the University of Tokyo, the High Energy Accelerator Research Organization (KEK), the Academia Sinica Institute for Astronomy and Astrophysics in Taiwan (ASIAA), and Princeton University. Funding was contributed by the FIRST program from Japanese Cabinet Office, the Ministry of Education, Culture, Sports, Science and Technology (MEXT), the Japan Society for the Promotion of Science (JSPS), Japan Science and Technology Agency (JST), the Toray Science Foundation, NAOJ, Kavli IPMU, KEK, ASIAA, and Princeton University. This paper makes use of software developed for the Large Synoptic Survey Telescope. We thank the LSST Project for making their code available as free software at  http://dm.lsst.org

The Pan-STARRS1 Surveys (PS1) have been made possible through contributions of the Institute for Astronomy, the University of Hawaii, the Pan-STARRS Project Office, the Max-Planck Society and its participating institutes, the Max Planck Institute for Astronomy, Heidelberg and the Max Planck Institute for Extraterrestrial Physics, Garching, The Johns Hopkins University, Durham University, the University of Edinburgh, Queen’s University Belfast, the Harvard-Smithsonian Center for Astrophysics, the Las Cumbres Observatory Global Telescope Network Incorporated, the National Central University of Taiwan, the Space Telescope Science Institute, the National Aeronautics and Space Administration under Grant No. NNX08AR22G issued through the Planetary Science Division of the NASA Science Mission Directorate, the National Science Foundation under Grant No. AST-1238877, the University of Maryland, and Eotvos Lorand University (ELTE) and the Los Alamos National Laboratory.

Based in part on data collected at the Subaru Telescope and retrieved from the HSC data archive system, which is operated by Subaru Telescope and Astronomy Data Center at National Astronomical Observatory of Japan.

GAMA is a joint European-Australasian project based around a spectroscopic campaign using the Anglo-Australian Telescope. The GAMA input catalogue is based on data taken from the Sloan Digital Sky Survey and the UKIRT Infrared Deep Sky Survey. Complementary imaging of the GAMA regions is being obtained by a number of independent survey programmes including GALEX MIS, VST KiDS, VISTA VIKING, WISE, Herschel-ATLAS, GMRT and ASKAP providing UV to radio coverage. GAMA is funded by the STFC (UK), the ARC (Australia), the AAO, and the participating institutions. The GAMA website is https://www.gama-survey.org/ .

This research used data obtained with the Dark Energy Spectroscopic Instrument (DESI). DESI construction and operations is managed by the Lawrence Berkeley National Laboratory. This material is based upon work supported by the U.S. Department of Energy, Office of Science, Office of High-Energy Physics, under Contract No. DE–AC02–05CH11231, and by the National Energy Research Scientific Computing Center, a DOE Office of Science User Facility under the same contract. Additional support for DESI was provided by the U.S. National Science Foundation (NSF), Division of Astronomical Sciences under Contract No. AST-0950945 to the NSF’s National Optical-Infrared Astronomy Research Laboratory; the Science and Technology Facilities Council of the United Kingdom; the Gordon and Betty Moore Foundation; the Heising-Simons Foundation; the French Alternative Energies and Atomic Energy Commission (CEA); the National Council of Humanities, Science and Technology of Mexico (CONAHCYT); the Ministry of Science and Innovation of Spain (MICINN), and by the DESI Member Institutions: www.desi.lbl.gov/collaborating-institutions. The DESI collaboration is honored to be permitted to conduct scientific research on I’oligam Du’ag (Kitt Peak), a mountain with particular significance to the Tohono O’odham Nation. Any opinions, findings, and conclusions or recommendations expressed in this material are those of the author(s) and do not necessarily reflect the views of the U.S. National Science Foundation, the U.S. Department of Energy, or any of the listed funding agencies.

\end{acknowledgments}

\section{Data availability}
All raw data related to the findings in the manuscript is publicly available and provided by the HSC survey\citep{2022_Li_xianchong}.

\appendix

\section{Cross-component of shear profiles}\label{app:x-shear}
\begin{figure*}
    \centering
    \includegraphics[width=\textwidth]{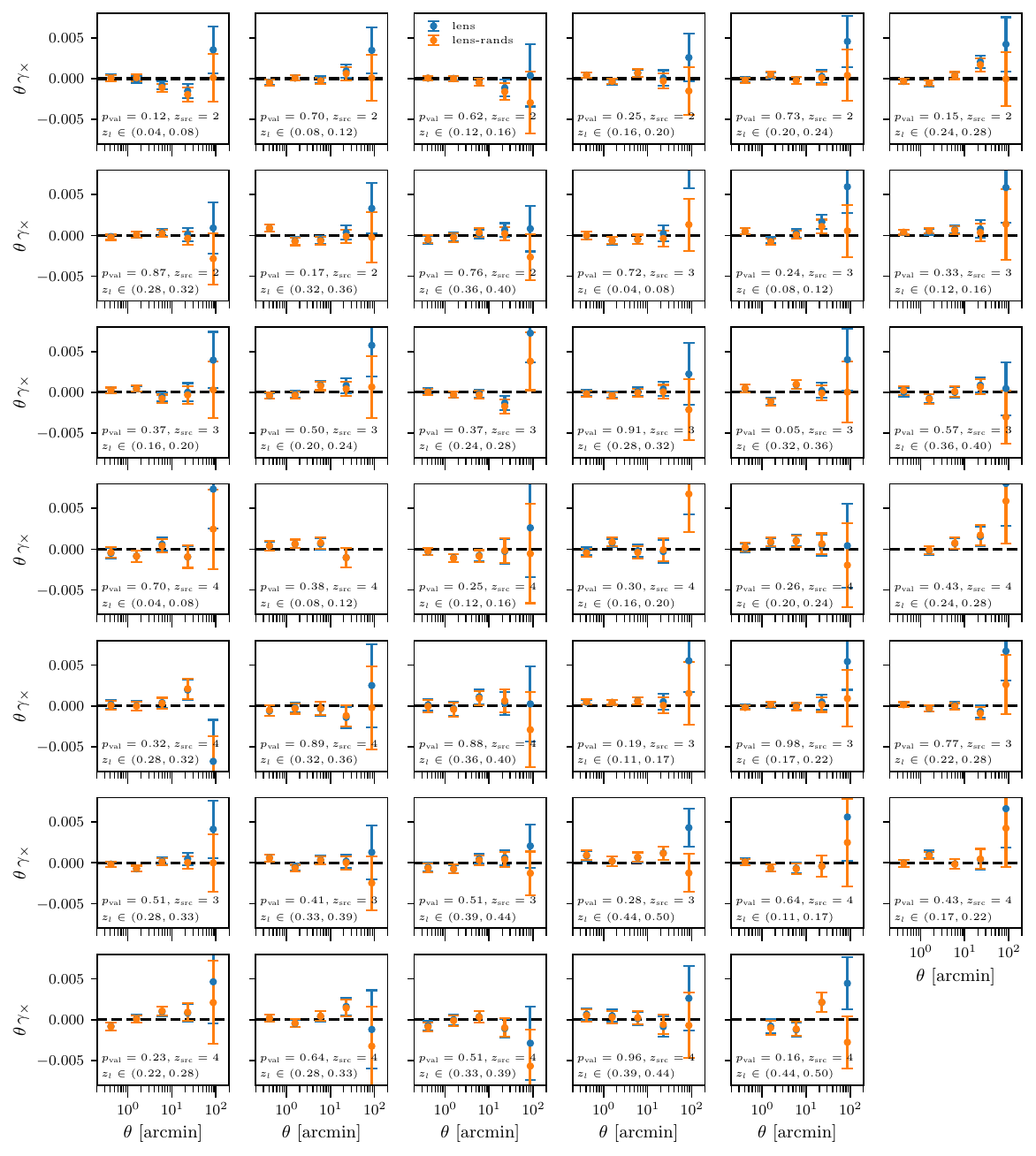}
    \caption{\textit{Cross shear measurements:} We show the cross component of the shear signal for each lens redshift bin \(z_l\) and source redshift bin \(z_{\rm src}\) combination. The blue data points, shown with shape noise error bars, represent the cross shear measured around the lenses. The orange data points indicate the signal after subtracting the measurements around random points to mitigate residual PSF systematics. The black dashed line denotes the expected null signal. In each panel, we report the \(p\)-value (\(p_{\rm val}\)) calculated from the random-subtracted measurements to quantify consistency with the null hypothesis.
}
    \label{fig:shear-cross-meas}
\end{figure*}

We perform a systematics check on the angular scales used in our tangential shear measurements, as shown in Fig.~\ref{fig:shear-meas}. Specifically, we measure the cross component of the shear around the lens galaxies and subtract the corresponding signal measured around random points to remove residual PSF-related systematics. In Fig.~\ref{fig:shear-cross-meas}, each panel shows the cross shear measurement for a given lens and source redshift bin. The blue data points with shape noise errors represent the cross shear measured around lenses, while the orange points show the signal after subtracting the measurements around random points. The black dashed line indicates the expected null signal.

For each lens--source bin combination, we compute a \(p\)-value ($p_{\rm val}$) for the null hypothesis using the random-subtracted measurements, shown in each plot. We find good agreement with the null, with \(p\)-values no smaller than 12 percent across all lens redshift bins, indicating no significant detection of systematic contamination. This supports the robustness of our tangential shear measurements across the full range of angular scales.

\section{Comparison using DESI-DR1}\label{app:comp-DESI-DR1}

We reran our analysis using galaxies from the DESI-DR1 dataset as lenses by applying the same lens selection criteria as listed in Table~\ref{tab:lenses} of our fiducial analysis. We measured the tangential shear signal around the DESI-DR1 lenses and assessed systematic effects similar to the fiducial analysis. We found no significant systematics within the angular scales used in our analysis. From these measurements, we inferred the corresponding shear-ratio signal, obtaining a total signal-to-noise ratio of 212.4. As in the fiducial analysis, we used these measurements jointly with those from BOSS DR12 to constrain photometric redshift (photo-\(z\)) bias parameters. We also combined them with HSC-Y3 cosmic shear measurements to derive cosmological constraints.

Our modeling of the shear-ratio measurements yields a best-fit \(\chi^2 = 147.12\) with 116.6 effective degrees of freedom. Similarly, the joint cosmological analysis results in a best-fit \(\chi^2 = 296\) with 250.9 effective degrees of freedom, indicating good agreement between the model and the data.

\begin{figure}
    \centering
    \includegraphics[width=\columnwidth]{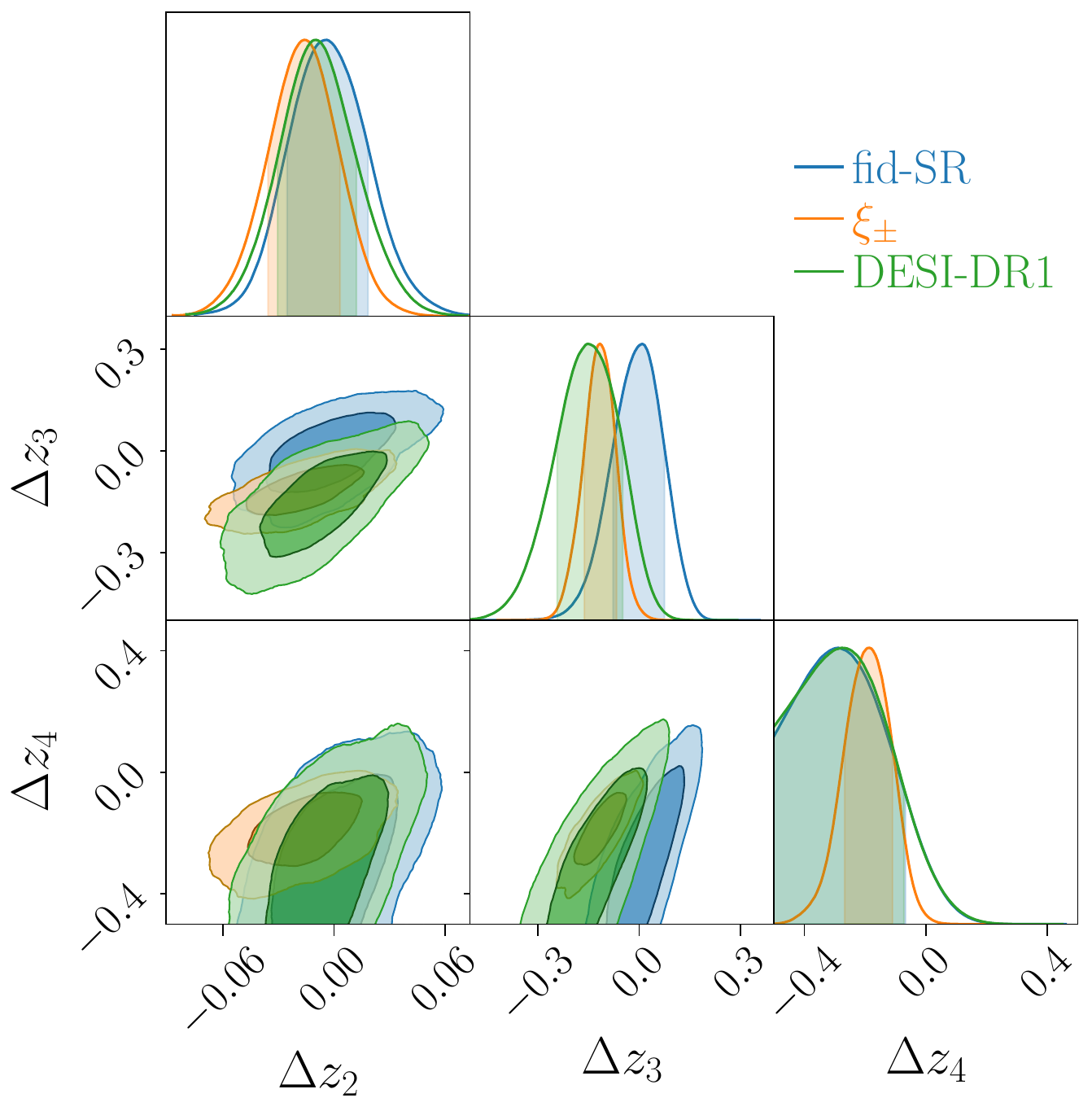}
    \caption{\textit{DESI-DR1 photo-\(z\) biases:} Comparison of constraints on the source redshift shift parameters \(\Delta z\). The blue and green having $68$ and $95$ confidence contours each, along with their corresponding posteriors, represent the constraints obtained from shear-ratio measurements using lenses from the fiducial sample and DESI-DR1, respectively. The orange contours and posterior show the constraints derived using only the HSC-Y3 cosmic shear measurements.}
    \label{Appendix:fig:photo-z-biases}
\end{figure}

\begin{figure}
    \centering
    \includegraphics[width=\columnwidth]{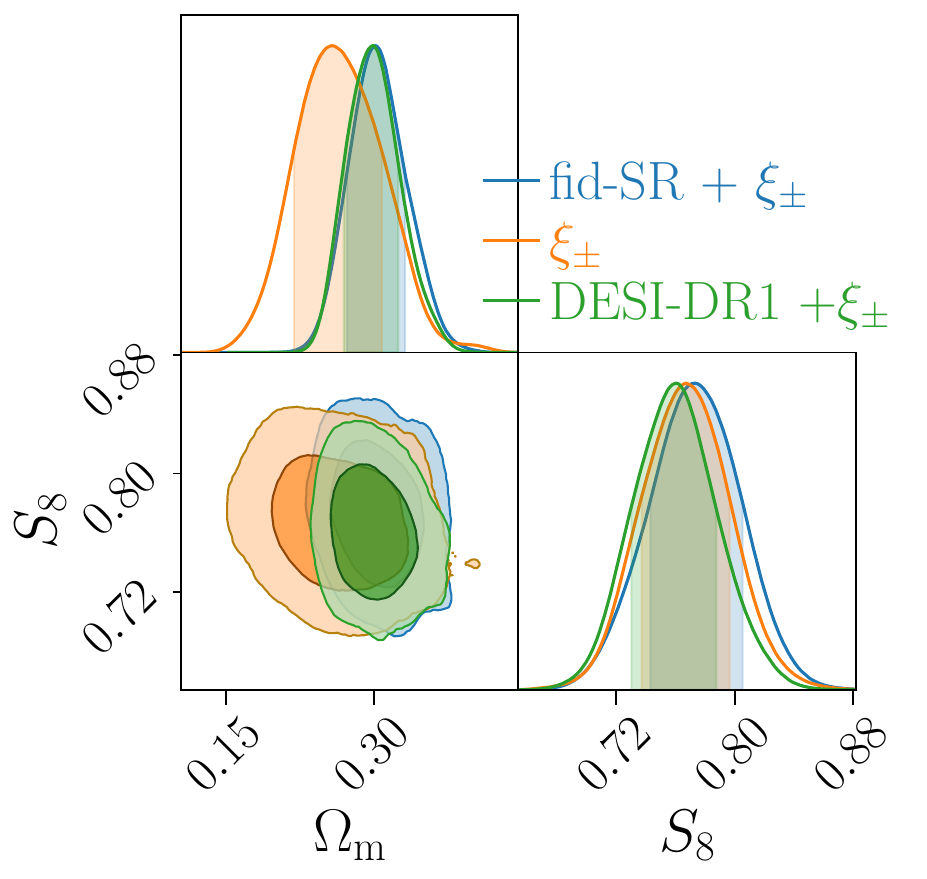}
    \caption{\textit{DESI-DR1 cosmological constraints:} Comparison of cosmological constraints from the combined analysis of HSC-Y3 cosmic shear with shear-ratio measurements. The blue and green contours show the combined constraints using the fiducial and DESI-DR1 lens samples, respectively. The orange contours represent the constraints derived from modeling the HSC-Y3 cosmic shear measurements alone.The contours denote the 68\% and 95\% confidence intervals.}

    \label{Appendix:fig:cosmology-constraints}
\end{figure}

In Figs.~\ref{Appendix:fig:photo-z-biases} and \ref{Appendix:fig:cosmology-constraints}, the green contours show our constraints derived using DESI-DR1 lenses, compared against those from the fiducial analysis and the HSC-Y3 cosmic shear results. Using DESI-DR1, we obtain the following constraints on the photometric redshift bias parameters: \(\Delta z_2 = -0.009_{-0.02}^{+0.022}\), \(\Delta z_3 = -0.151_{-0.099}^{+0.094}\), and \(\Delta z_4 = -0.35_{-0.275}^{+0.203}\). The corresponding cosmological constraints after combining with cosmic shear measurements are \(\Omega_{\rm m} = 0.299_{-0.027}^{+0.028}\) and \(S_8 = 0.76_{-0.029}^{+0.029}\). These results are in good agreement with the fiducial analysis for both photo-\(z\) bias and cosmological parameters.

However, we observe a small shift in the \(\Delta z_3\)--\(\Delta z_4\) contours compared to the fiducial analysis. We attribute this shift to the heterogeneous nature of the fiducial lens sample, which is heavily influenced by the GAMA survey. To test this, we reanalyzed the fiducial sample after removing all lenses within the overlapping GAMA footprint. Interestingly, this resulted in a shift in the \(\Delta z_3\)--\(\Delta z_4\) contour in the opposite direction to that seen in the cosmic shear analysis. This behavior suggests that such systematic shifts tend to average out when a homogeneous lens sample, such as DESI-DR1, is used.

\bibliographystyle{apsrev}
\bibliography{apssamp}

\end{document}